\documentclass[12pt]{article}
\usepackage{amssymb}
\usepackage{epsfig}
\voffset= - 1.0in
\hoffset= - 0.6in         

\catcode`\@=11
\def\marginnote#1{}

\newcount\hour
\newcount\minute
\newtoks\amorpm
\hour=\time\divide\hour by60
\minute=\time{\multiply\hour by60 \global\advance\minute by-\hour}
\edef\standardtime{{\ifnum\hour<12 \global\amorpm={am}%
        \else\global\amorpm={pm}\advance\hour by-12 \fi
        \ifnum\hour=0 \hour=12 \fi
        \number\hour:\ifnum\minute<10 0\fi\number\minute\the\amorpm}}
\edef\militarytime{\number\hour:\ifnum\minute<10 0\fi\number\minute}

\def\draftlabel#1{{\@bsphack\if@filesw {\let\thepage\relax
   \xdef\@gtempa{\write\@auxout{\string
      \newlabel{#1}{{\@currentlabel}{\thepage}}}}}\@gtempa
   \if@nobreak \ifvmode\nobreak\fi\fi\fi\@esphack}
        \gdef\@eqnlabel{#1}}
\def\@eqnlabel{}
\def\@vacuum{}
\def\draftmarginnote#1{\marginpar{\raggedright\scriptsize\tt#1}}

\def\draft{\oddsidemargin -.5truein
        \def\@oddfoot{\sl preliminary draft \hfil
        \rm\thepage\hfil\sl\today\quad\militarytime}
        \let\@evenfoot\@oddfoot \overfullrule 3pt
        \let\label=\draftlabel
        \let\marginnote=\draftmarginnote
   \def\@eqnnum{(\theequation)\rlap{\kern\marginparsep\tt\@eqnlabel}%
\global\let\@eqnlabel\@vacuum}  }

\def\appname{Appendix}
\newcounter{app}
\def\theapp{\Alph{app}}
\def\app{\par
   \addvspace{4ex}
   \@afterindentfalse
  \secdef\@app\@dapp}
\def\@app[#1]#2{\ifnum \c@secnumdepth >\m@ne
        \refstepcounter{app}
        \addcontentsline{toc}{app}{\theapp
        \hspace{1em}#1}\else
      \addcontentsline{toc}{app}{ #1}\fi
   {\parindent \z@ \raggedright
    \Large \bf \appname~\theapp .
   \Large  \bf 
    #2}\nobreak
   \vskip 4ex   \noindent
\setcounter{equation}{0}
\def\theequation{\Alph{app}.\arabic{equation}}}
\def\@dapp#1{%
{\parindent \z@ \raggedright  \bf #1}\par\nobreak}
\def\l@app#1#2{\addpenalty{\@secpenalty}%
   \addvspace{1em plus\p@}%
   \begingroup
   \@tempdima 3em
     \parindent \z@ \rightskip \@pnumwidth
     \parfillskip -\@pnumwidth
     { \bf
     \leavevmode
     #1\hfil \hbox to\@pnumwidth{\hss #2}}\par
     \nobreak
   \endgroup}

\def\d{\partial}

\def\bea{\begin{eqnarray}}
\def\eea{\end{eqnarray}}

\def\beq{\begin{equation}}
\def\eeq{\end{equation}}
\def\ba{\beq\new\begin{array}{c}}
\def\ea{\end{array}\eeq}
\def\be{\ba}
\def\ee{\ea}
\def\stackreb#1#2{\mathrel{\mathop{#2}\limits_{#1}}}
\def\Tr{{\rm Tr}}

\makeatletter
\newdimen\normalarrayskip              
\newdimen\minarrayskip                 
\normalarrayskip\baselineskip
\minarrayskip\jot
\newif\ifold             \oldtrue            \def\new{\oldfalse}
\def\arraymode{\ifold\relax\else\displaystyle\fi} 
\def\eqnumphantom{\phantom{(\theequation)}}     
\def\@arrayskip{\ifold\baselineskip\z@\lineskip\z@
     \else
     \baselineskip\minarrayskip\lineskip2\minarrayskip\fi}
\def\@arrayclassz{\ifcase \@lastchclass \@acolampacol \or
\@ampacol \or \or \or \@addamp \or
   \@acolampacol \or \@firstampfalse \@acol \fi
\edef\@preamble{\@preamble
  \ifcase \@chnum
     \hfil$\relax\arraymode\@sharp$\hfil
     \or $\relax\arraymode\@sharp$\hfil
     \or \hfil$\relax\arraymode\@sharp$\fi}}
\def\@array[#1]#2{\setbox\@arstrutbox=\hbox{\vrule
     height\arraystretch \ht\strutbox
     depth\arraystretch \dp\strutbox
     width\z@}\@mkpream{#2}\edef\@preamble{\halign
\noexpand\@halignto
\bgroup \tabskip\z@ \@arstrut \@preamble \tabskip\z@ \cr}%
\let\@startpbox\@@startpbox \let\@endpbox\@@endpbox
  \if #1t\vtop \else \if#1b\vbox \else \vcenter \fi\fi
  \bgroup \let\par\relax
  \let\@sharp##\let\protect\relax
  \@arrayskip\@preamble}
\def\eqnarray{\stepcounter{equation}%
              \let\@currentlabel=\theequation
              \global\@eqnswtrue
              \global\@eqcnt\z@
              \tabskip\@centering
              \let\\=\@eqncr
 \halign to \displaywidth\bgroup
    \eqnumphantom\@eqnsel\hskip\@centering
    $\displaystyle \tabskip\z@ {##}$%
    \global\@eqcnt\@ne \hskip 2\arraycolsep
         $\displaystyle\arraymode{##}$\hfil
    \global\@eqcnt\tw@ \hskip 2\arraycolsep
         $\displaystyle\tabskip\z@{##}$\hfil
         \tabskip\@centering
    &{##}\tabskip\z@\cr}
\begingroup\ifx\undefined\newsymbol \else\def\input#1 {\endgroup}\fi
\newfont{\hr}{msbm10}
\newfont{\ams}{msam10}

\font\numbers=cmss12
\font\upright=cmu10 scaled\magstep1
\def\stroke{\vrule height8pt width0.4pt depth-0.1pt}
\def\topfleck{\vrule height8pt width0.5pt depth-5.9pt}
\def\botfleck{\vrule height2pt width0.5pt depth0.1pt}
\def\Zmath{\vcenter{\hbox{\numbers\rlap{\rlap{Z}\kern 0.8pt\topfleck}\kern
2.2pt
                   \rlap Z\kern 6pt\botfleck\kern 1pt}}}
\def\Qmath{\vcenter{\hbox{\upright\rlap{\rlap{Q}\kern
                   3.8pt\stroke}\phantom{Q}}}}
\def\Nmath{\vcenter{\hbox{\upright\rlap{I}\kern 1.7pt N}}}
\def\Cmath{\vcenter{\hbox{\upright\rlap{\rlap{C}\kern
                   3.8pt\stroke}\phantom{C}}}}
\def\Rmath{\vcenter{\hbox{\upright\rlap{I}\kern 1.7pt R}}}
\def\Z{\ifmmode\Zmath\else$\Zmath$\fi}
\def\Q{\ifmmode\Qmath\else$\Qmath$\fi}
\def\N{\ifmmode\Nmath\else$\Nmath$\fi}
\def\C{\ifmmode\Cmath\else$\Cmath$\fi}
\def\R{\ifmmode\Rmath\else$\Rmath$\fi}

\def\stackreb#1#2{\mathrel{\mathop{#2}\limits_{#1}}}
\def\Tr{{\rm Tr}}
\def\res{{\rm res}}
\def\Bf#1{\mbox{\boldmath $#1$}}
\def\balpha{{\Bf\alpha}}

\def\bmu{{\Bf\mu}}

\def\pt{\partial}
\def\d{\partial}

\def\Im{{\rm Im}}
\def\Re{{\rm Re}}

\def\half{{\textstyle{1\over2}}}
\def\2{{1\over 2}}
\def\ntwo{${\mathcal N}=2\;$}

\def\none{${\mathcal N}=1\;$}
\def\ntwot{${\mathcal N}=(2,2)\;$}

\def\beq{\begin{equation}}
\def\eeq{\end{equation}}
\def\ba{\beq\new\begin{array}{c}}
\def\ea{\end{array}\eeq}
\def\be{\ba}
\def\ee{\ea}
\def\stackreb#1#2{\mathrel{\mathop{#2}\limits_{#1}}}
\def\theequation{\thesection.\arabic{equation}}

\newcommand{\rf}[1]{(\ref{#1})}

\begin{document}


\begin{flushright}
FIAN/TD-19/09\\
ITEP/TH-46/09\\
YITP\ -\ 09\ -\ 106\\
FTPI-MINN-09/46\\
UMN-TH-2829/09
\end{flushright}

\vspace*{0.5cm}

\begin{center}
\baselineskip20pt
{\bf \LARGE Strong versus Weak Coupling Confinement\\
\vspace{0.3cm}
in \ntwo Supersymmetric QCD}
\end{center}
\bigskip
\begin{center}
{\bf\large A.~Marshakov}$^{a,b}$ {\large and}
{\bf\large A.~Yung}$^{c,d}$

\end {center}
\begin{center}

$^a${\it Theory Department, P.~N.~Lebedev Physics Institute and\\
Institute for Theoretical and Experimental Physics,
Moscow, Russia}\\
$^b${\it Yukawa Institute for Theoretical Physics,
Kyoto, Japan}\\
$^c${\it Petersburg Nuclear Physics Institute, Gatchina, Russia}\\
$^d${\it  William I. Fine Theoretical Physics Institute,
University of Minnesota,\\
Minneapolis, USA}\\

\bigskip\bigskip\medskip

\begin{center}
{\large\bf Abstract} \vspace*{.2cm}
\end{center}

\begin{quotation}
We consider \ntwo supersymmetric QCD with the gauge group
$SU(N_c)=SU(N+1)$ and $N_f$ number of quark matter multiplets, being
perturbed by a small mass term for the adjoint matter, so that its
Coulomb branch shrinks to a number of isolated vacua. We discuss the
vacuum where $r=N$ quarks develop VEV's for $N_f\geq 2N=2N_c-2$
(in particular, we focus on the $N_f= 2N$ and $N_f= 2N+1$ cases).
 In the equal quark mass limit at
large masses this vacuum stays at weak coupling, the low-energy theory
has $U(N)$ gauge symmetry and one
observes the non-Abelian confinement of monopoles. As we reduce the
average quark mass and enter the strong coupling regime the quark
condensate transforms into the condensate of dyons. We show that the
low energy description in the strongly-coupled domain for the
original theory is given by $U(N)$ dual gauge theory of $N_f\geq 2N$
light non-Abelian dyons, where the condensed dyons still cause the
confinement of monopoles, and not of the quarks, as can be thought
by naive duality.

\end{quotation}
\end{center}


\newpage

\setcounter{footnote}{0}
\setcounter{equation}{0}

\section{Introduction}

In this paper we consider \ntwo supersymmetric QCD with the gauge
group $SU(N_c)=SU(N+1)$ and $N_f$ fundamental quark matter
multiplets. This \ntwo theory is perturbed by a small mass term
$\mu\Tr\Phi^2$ for adjoint matter, so that the Coulomb branch
shrinks to a number of isolated vacua, whose study was initiated in
\cite{MY}. We continue the program of studying the features of confinement in
\ntwo supersymmetric QCD in the regime, when one can trust the
semiclassical string solutions. Mostly interesting part of this
study concerns the case, when on the intermediate scales $\sqrt{\mu
m}$ (for $\mu\ll m$, the average quark mass) the non-Abelian $SU(N)$
symmetry is restored and one may think of a {\em non-Abelian}
confinement. In
\cite{MY} we have found the confinement of monopoles with the
non-Abelian spectrum at weak coupling, and a natural next question
is what happens to this picture, if one manages to move it into the
strong coupling regime of the original theory. For that purpose we
need to use the details of the Seiberg-Witten exact solution
\cite{SW1,SW2}.

In order to achieve this aim, we start with the vacuum,  where
maximum number of quarks condense in general position. If $N_f\geq
2N$ (in particular, we focus on the cases $N_f= 2N=2N_c-2$ and $N_f=
2N+1=2N_c-1$) this vacuum (where $N$ quarks have nonvanishing
condensates) stays at weak coupling, and (in the equal quark mass
limit) at low energies the theory in the vicinity of such vacuum has
a non-Abelian description in terms of the
non-asymptotically free $SU(N)\times U(1) \simeq
U(N)$ gauge theory
\cite{APS,CKM,MY} with $N_f$ light quark flavors. Condensation of
quarks ensures the monopole confinement: such theory supports the
non-Abelian strings \cite{HT1,ABEKY,SYmon,HT2} (see also reviews
\cite{SYrev,Trev,Jrev,Trev2}), which confine monopoles. We address the following
question: what happen to the confinement when we change parameters
of the theory and go to the strong coupling regime.

We shall see, that as we reduce the average quark mass and enter the
strong coupling domain in the "mass-plane", the original quarks
change their quantum numbers and transform into dyons. We show that
the low energy effective theory is given by the $U(N)$ dual gauge
theory with $N_f\geq 2N$ light non-Abelian dyons. The dual theory at
strong coupling also supports the non-Abelian strings, much as the
original one. However, we find that these strings in the dyonic
condensate still confine monopoles! Thus, in contrast to naive
expectations, one still get a confinement of monopoles (not quarks!)
at the dual strong-coupling regime of
\ntwo supersymmetric QCD.

This picture can be compared with studied recently in \cite{SYdual,SYcross}
for \ntwo supersymmetric QCD with the gauge group $U(N)$ and $N\leq
N_f<2N$ flavors. The parameter interpolating between weak and strong
coupling regimes is the coefficient of the Fayet-Iliopoulos (FI)
term $\xi$, which plays the same role as $\sqrt{\mu m}$ in our
setup. At large values of $\xi$ the theory with $N\leq N_f$ is in
weak-coupled phase, while at small $\xi$ this theory goes into the
strong coupling regime. It has been shown in
\cite{SYdual,SYcross}, that the theory upon reducing $\xi$ exhibits a
crossover transition, and below crossover it is described in terms
of dual $U(N_f-N)$ theory of $N_f$ dyons, similar to the Seiberg
duality \cite{Sdual,IS} in \none supersymmetric QCD, where emergence
of the dual gauge group $U(N_f-N)$ was first recognized, see also
\cite{APS}.

Although the transition in the full theory is smooth the low energy
effective descriptions in two regimes differ drastically: the gauge
groups and spectra of light states are different, and the
perturbative and non-perturbative states (mesons formed by
monopole-dyon pairs connected by confining strings) interchange upon
passing from one regime to another. Moreover, it is shown in
\cite{SYdual,SYcross} that in the strong coupling region at small
$\xi$, the states confined by non-Abelian strings are still
monopoles, which is very close to the conclusions of present paper.
However, below we shall see, that the low energy effective theories
for $N_f\geq 2N=2N_c-2$ are essentially different from the $N_f <2N$
case. We show that in our case with large number of flavors, the low
energy description is still given by the $U(N)$ dual  gauge theory
with $N_f\geq 2N$ light non-Abelian dyons, and there is no crossover,
in contrast to the case
$N_f< 2N$, when the dual gauge group becomes $U(N_f-N)$.

The paper is organized as follows. In sect.~\ref{ss:class} we
present our theory and in sect.~\ref{ss:semi} study, what happens
with the quark $r=2$ vacuum in the weak coupling regime - at large
$m$ - semiclassically. Mostly in this paper we consider the simplest
but mostly illustrative case of $N_c=3$ or $N=2$, so that the gauge
group in ultraviolet is $SU(3)$. In sect.~\ref{ss:exact} we use the
Seiberg-Witten curves and differentials to study our theory at
strong coupling at small $m$. In sect.~\ref{ss:dualact} we present
the   action of the dual theory, describing effectively the low
energy limit of our theory  at small $m$ and, finally, in
sect.~\ref{ss:conf} we calculate the fluxes of strings leading
confinement of monopoles in dual theory. Sect.~\ref{ss:concl}
contains our conclusions, and some details are collected in Appendices.

\section{Classical theory
\label{ss:class}}
\setcounter{equation}{0}

\subsection{Supersymmetric QCD
\label{ss:two}}

The particular theory we are going to consider in this paper is \ntwo supersymmetric
QCD with the gauge group $SU(N_c)=SU(3)$ and $N_f=4,5$ flavors of
fundamental matter hypermultiplets (quarks). Generically we choose
different values for the quark multiplets masses, $m_A\neq m_B$, $A,B=1,\ldots,N_f$.
However, our final goal is to consider the case
when at least some of quark masses coincide $\Delta m_{AB}\equiv m_A-m_B\to 0$.
In this limit the global symmetry $SU(N)_{C+F}=SU(2)_{C+F}$ emerges, which is responsible in particular
for presence of the nonabelian strings and the nonabelian features of confinement.

The bosonic part of the action of our model reads
\be
\label{qed}
S = \int d^4x \left[\frac1{4g^2}
\Tr\left(F_{\mu\nu}\right)^2
+
\frac1{g^2}\Tr\left(D_{\mu}\Phi D_\mu\Phi^\dagger\right)  \right. +
\\
 + \left. \left|\nabla_{\mu}
Q_{A}\right|^2 + \left|\nabla_{\mu} {\tilde{Q}}^{A}\right|^2
+ V(Q,\tilde{Q},\Phi)\right]
\ee
Here $D_{\mu}$ is the covariant derivative in the adjoint representation
of  $SU(N_c)$, while
\be
\label{nabsun}
\nabla_\mu=\partial_\mu-i A_{\mu}=\partial_\mu-i A_{\mu}^aT^a
\ee
is covariant derivative
in the fundamental representation. Normally we
suppress the color matrix indices, they can be restored, for example, in
the $SU(3)$ case using the Gell-Mann matrices $T^a=\| (T^a)^k_l\|$, $a=1,\ldots,8$ and $k,l=1,2,3$
with the commutator  $[T^a,T^b]=f^{abc}T^c$ and normalized as $\Tr (T^a T^b)=\half\delta^{ab}$.
The adjoint scalar  $\Phi = \sqrt{2}a^aT^a$ is a superpartner of the gauge field in the vector
supermultiplet.
The scalar components of quark hypermultiplets $Q^k_A$ and
$\tilde{Q}_k^{A}$ are
$N_c\times N_f = 3\times N_f $ matrices with color ($k=1,\ldots,N_c$) and flavor
($A=1,\ldots,N_f$) indices.

The potential $V(Q,\tilde{Q},\Phi)$ in the Lagrangian (\ref{qed})
is a sum of the FI $D$- (the first line) and $F$- (the second line) terms,
\be
V(Q,\tilde{Q},\Phi) =
 \frac{g^2}{2}
\left( \bar{Q}^AT^aQ_A - \tilde{Q}^A T^a\bar{\tilde{Q}}_A +
\frac{1}{g^2}\,f^{abc} \bar a^b a^c \right)^2 +
\\
+ g^2\left| \sqrt{2}\tilde{Q}^A T^a Q_A +\mu a^a\right|^2+
\sum_{A=1}^{N_f} \left(\left|(\Phi+m_A)Q^A\right|^2+
\left|\tilde{Q}_A(\Phi+m_A)\right|^2\right)
\label{pot}
\ee
where the sum over repeated flavor indices $A$ is implied.

The potential \rf{pot} implies, that the original $SU(3)$ theory is
perturbed by adding a small
mass term for the adjoint matter,  via the superpotential
${\cal W}=\mu{\rm Tr}\Phi^2$.
Generally speaking, this superpotential breaks
\ntwo supersymmetry down to \none, and the Coulomb branch shrinks to
a number of  isolated \none vacua \cite{APS,CKM}.
In the limit of $\mu\to 0$ these vacua correspond to special
singular points on the Coulomb branch, when a pair of monopoles/dyons or
quarks become massless. $N_c=3$ of them (often referred to as the
Seiberg-Witten vacua) are always at strong coupling; these vacua also
exist in \none pure $SU(N_c)$ gauge theory.

There are also vacua of a different type,
to be referred to as quark vacua, which may or may not be at weak
coupling, dependently on the values of the quark masses  $m_A$, for
$m_A\gg \Lambda_{{\rm SU}(3)}$ quark vacua with $\langle Q^A\rangle \neq 0$
are in the weak coupling regime.
These vacua are characterized by an integer $r$, counting the
number of condensed flavors, and the number of gauge non-equivalent ones
equals to $(N_c-r)C^{N_f}_r$ \cite{APS,CKM,MY}.
For the $SU(3)$ gauge theory one has therefore
$2N_f$ vacua with $r=1$ and $N_f(N_f-1)/2$ with $r=2$.

Below we concentrate mostly on physics in $r=2$ vacua.
In particular, for these vacua we have the phenomenon
of restoration of global $SU(2)_{C+F}$ symmetry
if two masses of the condensed quarks coincides, when
the $\mathbb{Z}_2$-strings develop orientational
zero modes and therefore are called non-Abelian
\cite{HT1,ABEKY,SYmon,HT2}.


\subsection{Quark vacua}


Consider $r=2$ vacuum with nonvanishing VEV's $\langle Q^A\rangle \neq 0$ of, say,
$A=1,2$ flavors (12-vacuum). At large non-degenerate values of masses $m_A$, $A=1,2$,
this vacuum is in weak coupling and semiclassical analysis
is applicable. The adjoint scalar develop the following VEV's
(see \cite{MY} for more details):
\be
\label{fir2}
\Phi = - \left(
\begin{array}{ccc}
  m_1 &  &  \\
   & m_2 &  \\
   &  & -m_1-m_2
\end{array}\right)
\ee The same can be rewritten in $T^a$-components as
\be
\sqrt{2}\langle a^3\rangle = \langle \balpha_{12}\cdot {\bf a}
\rangle = -\Delta m, \qquad \langle a^8\rangle = - \sqrt{3\over
2}\langle\bmu_{12}\cdot{\bf a}\rangle = - \sqrt{6}m
\label{avev}
\ee
where we have introduced \be m=\half (m_1+m_2), \qquad \Delta
m=m_1-m_2 \ee By gauge rotations the  VEV's of two quark flavors can
be chosen as
\be
\label{q2}
Q_A^k = \delta_A^k \sqrt{\mu M_A} ,\ \ \ A=1,2,\ \ k=1,2
\\
M_1=2m_1+m_2,\ \ \ M_2=m_1+2m_2
\ee
\begin{figure}[tp]
\epsfysize=7cm
\centerline{\epsfbox{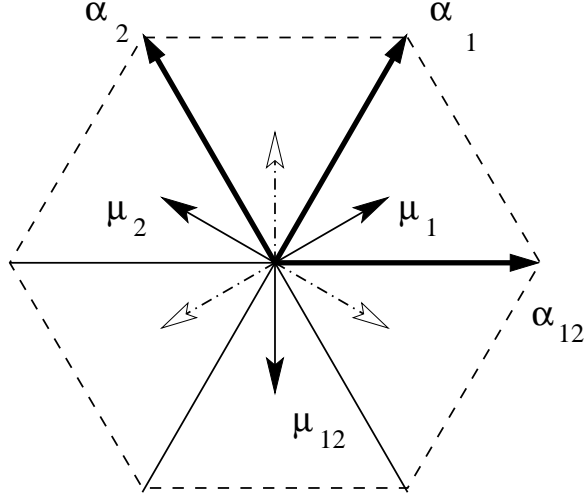}}
\caption{Roots and fundamental weights for the $SU(3)$ gauge group in its Cartan plane.
The roots are canonically normalized, as $\balpha^2=2$. The simple roots are $\balpha_{12}=
\balpha_1-\balpha_2$ and $\balpha_2$, while $\balpha_1=\balpha_{12}+\balpha_2$ is the
``highest'' root. The notations are chosen for the roots to be orthogonal
$\balpha_i\bmu_j=\delta_{ij}$, $i,j=1,2$ to the weights $\bmu_1$ and $\bmu_2$ of the
fundamental representations ${\bf 3}$ (the weights of the dual fundamental
representation ${\bar{\bf 3}}$ are depicted with dashed lines).}
\label{fi:su3}
\end{figure}
In the limit $\mu\to 0$ we move towards Coulomb branch where
two quarks become massless. The conditions of their
masslessness follow from \rf{pot}, (\ref{avev}). They read
\be
a_{A}+m_A=0,\ \ \ A=1,2
\label{12qmass}
\ee
where
\be
a_{A}=\bmu_A\cdot{\bf a} = \left\{
\begin{array}{c}
  \frac{a^8}{\sqrt{6}}+\frac{a^3}{\sqrt{2}}\ \ \ A=1 \\
  \frac{a^8}{\sqrt{6}}-\frac{a^3}{\sqrt{2}}\ \ \ A=2
\end{array}\right.
\label{aw}
\ee
In general the mass of a BPS state is given by Seiberg-Witten
formula \cite{SW1,SW2}
\be
m_{({\bf q}_e,{\bf q}_m)}=\left|\,\bf{q}_e\cdot\bf{a}+
\bf{q}_m\cdot\bf{a}_D +B_A m_A \right|
\label{BPSmass}
\ee
where ${\bf a}$ and ${\bf a}_D$ are two vectors in the Cartan plane
of $SU(3)$ gauge group with Cartesian components $(a^3,a^8)$,
$(a^3_D,a^8_D)$ respectively, while electric and magnetic charges of
a BPS state are given by the vectors ${\bf q}_e$ and ${\bf q}_m$
from the (dual) Cartan plane, see fig.~\ref{fi:su3}. The flavor
charge ${\bf B}=\{B_A\}$ is a $N_f$-vector determined by the
transformation properties of a state with respect to the global
flavor group $U(N_f)$ broken down to $U(1)^{N_f}$ in the case of
generic masses $m_A$  (for example, the flavor charge of the first
quark is ${\bf B}_1=(1,0,\ldots)$, the second quark is ${\bf
B}_2=(0,1,0,\ldots)$, etc). From (\ref{q2}) it follows that massless
quarks in the 12-vacuum have electric charges given by two weight
vectors $\bmu_1$ and $\bmu_2$, indeed, using the mass formula
(\ref{BPSmass}), one easily find that their masses
$m^A_{\bmu_A}=|\bmu_A\cdot{\bf a}+m_A|=0$ for both $A=1,2$ due to
(\ref{12qmass}), \rf{aw}. Note also, that  quarks with same color
charges but different flavors have masses $\Delta m$, e.g.
\be
m_{\bmu_1}^{A=2}=|\bmu_1\cdot{\bf a}+m_2|=|\Delta m|=|\bmu_2\cdot{\bf a}+m_1|=
m_{\bmu_2}^{A=1}
\label{w1221}
\ee
From (\ref{BPSmass}), (\ref{avev}) one also gets
\be
m_{\balpha_{12}}=\left|\balpha_{12}\cdot{\bf a}\right|=
\sqrt{2}\left|a^3\right|=\left|\Delta m_{12}\right|
\label{Wmass}
\ee
for the W-boson of the $SU(2)$, generated by the root vector
$\balpha_{12}=\balpha_1-\balpha_2$.

In the special case, when $\Delta m_{12}=m_1-m_2=0$, the masses
\rf{w1221} and \rf{Wmass} vanish, and this is a sign of (partial) restoration of
the non-Abelian symmetry. Indeed, the $SU(3)$ gauge group is broken
now to $U(2) \simeq SU(2)\times U(1)$ at high scale $m=m_1=m_2$. In
the effective low-energy $U(2)$ theory at small adjoint mass $\mu$
there is a crucial simplification. Since the chiral superfield
${\cal A}=a +\sqrt{2}\lambda\theta +F_a\theta^2$, the ${\cal N}=2$
superpartner of the $U(1)$ gauge field (embedded in the
$T^8$-direction into the $SU(3)$ gauge group), it not charged under
the gauge group, the superpotential ${\cal W}=\mu{\rm Tr}\Phi^2$ can
be truncated into the linear superpotential ${\cal W}_{{\cal A}}
\sim\xi{\cal A}$, with
\be
\xi = 6\mu m
\label{defxi}
\ee
which does {\em not} break \ntwo supersymmetry
\cite{HSZ,VY}. Of course, keeping higher order terms in $\mu{\rm Tr}\Phi^2$
would inevitably explicitly break \ntwo SUSY.

The bosonic part of the low energy effective action of the $U(2)$
theory reads
\be
S=\int d^4x \left[\frac1{4g^2_2}
\left(F^{\alpha}_{\mu\nu}\right)^2 +
\frac1{4g^2_1}\left(F^{8}_{\mu\nu}\right)^2
+
\frac1{g^2_2}\left|D_{\mu}a^\alpha\right|^2 +\frac1{g^2_1}
\left|\partial_{\mu}a^8\right|^2 \right.
\\
+\left. \left|\nabla_{\mu} Q_A\right|^2 + \left|\nabla_{\mu}
\bar{\tilde{Q}}^{A}\right|^2 +V(Q_A,\tilde{Q}^A,a^\alpha,a^8)\right]
\label{u2model}
\ee
Here $\alpha=1,2,3$, and $D_{\mu}$ is the covariant derivative in the adjoint representation
of  $SU(2)$, while
\be
\nabla_\mu=\partial_\mu -\frac{i}{2\sqrt{3}}\; A_{\mu}
-i A^{\alpha}_{\mu}\, \frac{\tau^\alpha}{2}
\label{nabsu2}
\ee
with the Pauli $\tau^\alpha$-matrices, normalized as
$\Tr(\tau^\alpha\tau^\beta)=2\delta_{\alpha\beta}$. The coupling
constants $g_1$ and $g_2$ correspond to the $U(1)$  and  $SU(2)$
sectors respectively, these couplings are equal at the scale $m$ of
breaking the $SU(3)$ symmetry down to $SU(2)\times U(1)$, but
generally split below this scale due to nontrivial renormalization
group flow in the $SU(2)$ sector (except for the conformal case with
$N_f=4$). The $U(1)$ charges $\pm1/2\sqrt{3}$ of the fundamental
matter fields are fixed by normalization of the $T^8$-generator in
the original $SU(3)$ gauge. Note that in the above action color
indices of quark fields $Q^k_A$ run over the restricted $SU(2)$
subset $k=1,2$.

The scalar potential $V(Q_A,\tilde{Q}^A,a^\alpha,a)$ in the action
(\ref{u2model}) has the form,
\be
V(Q_A,\tilde{Q}^A,a^\alpha,a) =
 \frac{g^2_2}{2}
\left(
 \bar{Q}^A\,\frac{\tau^\alpha}{2} Q_A -
\tilde{Q}^A \frac{\tau^\alpha}{2}\,\bar{\tilde{Q}}_A +
\frac{1}{g^2_2}\,  \varepsilon^{\alpha\beta\gamma} \bar a^\beta a^\gamma
 \right)^2 +
\\
+ \frac{g^2_1}{24}
\left(\bar{Q}^A Q_A - \tilde{Q}^A \bar{\tilde{Q}}_A \right)^2+
 2g^2_2\left| \tilde{Q}^A \frac{\tau^\alpha}{2} Q_A \right|^2+
\frac{g^2_1}{6}\left| \tilde{Q}^A Q_A  -\xi\right|^2 +
\\
+\sum_{A=1}^{N_f} \left[\left|\left(\frac{a^8}{\sqrt{6}}
+{\tau^\alpha a^\alpha\over\sqrt{2}} +m_A\right)Q_A\right|^2+
\left|\left(\frac{a^8}{\sqrt{6}}
+{\tau^\alpha a^\alpha\over\sqrt{2}}+ m_A\right)\bar{\tilde{Q}}_A
\right|^2 \right]
\label{u2pot}
\ee
The adjoint fields in 12-vacuum now develop the VEV's
\be
\langle a^3\rangle =0,\ \ \ \ \ \langle a^8\rangle =- \sqrt{6}m
\label{avevna}
\ee
being just particular case of \rf{avev} at $m=m_1=m_2$.
The quark's condensates in this case acquire the color-flavor locked form
\be
\langle Q^k_A\rangle =\langle \tilde{Q}_k^A\rangle =\sqrt{
\frac{\xi}{2}}\underbrace{ \left(
\begin{array}{ccccc}
1 & 0 & 0& \ldots & 0\\
0 & 1& 0& \ldots & 0\\
\end{array}\right)}_{N_f}
\label{qvev}
\ee
where we restrict the color and flavor indices to
$k=1,2$ and $A=1,\ldots,N_f$ and present the quark fields as
$2\times N_f$ matrices in the tensor product of color and flavor spaces.

These $r=2$ vacua exist in a theory with two flavors already, other
flavors do not play any role in the classical theory. Note however,
that if the number of flavors is not large enough (for example,
$N_f=2$) the non-Abelian nature of  low energy theory on the Coulomb
branch is ruined at $\Delta m=0$ by the strong coupling effects, and
the low energy theory becomes Abelian (for example, it is well known
that W-bosons decay due to the presence of curves of marginal
stability and the gauge sector contains only two photons). Hence, we
need to consider a theory with $N_f=4$ or $N_f=5$ in order to ensure
that $SU(2)\times U(1)$ low energy theory is conformal or infrared
free with $\beta_{SU(2)}\leq 0$, and does not run into strong
coupling\footnote{An alternative way is to consider the theory away
from the Coulomb
 branch at $\xi\gg\Lambda_{{\cal N}=2}$, see e.g. \cite{SYmon,SYrev}.}.
 In this case classically unbroken on the Coulomb branch
 (at $\mu=0$) gauge symmetry $SU(2)\times U(1)$ remains unbroken
 on the quantum level \cite{APS}.

The color-flavor locked form of the quark VEV's in (\ref{qvev}) and
the absence of VEV of the adjoint scalar $a^\alpha$ condensate
(\ref{avevna}) result in the fact that, while both gauge and flavor
$SU(2)$ subgroups are broken by the quark condensation, the diagonal
$SU(2)_{C+F}$ survives as a global symmetry. It is seen explicitly,
since all states in the low-energy theory combine into multiplets of
$SU(2)_{C+F}$. In particular, on Coulomb branch at $\mu=0$ the
``wrong flavor'' quarks (\ref{w1221}) become massless at $\Delta
m=0$ and are unified with the always massless in $r=2$ vacuum
``right flavor'' quarks (\ref{12qmass}) to form ${\bf 2}\otimes{\bf
2}={\bf 3}+{\bf 1}$ of $SU(2)_{C+F}$, while two massless photons
$A_{\mu}^8$ and $A_{\mu}^3$ are combined with $W^{\pm}$-bosons (with
vanishing masses (\ref{Wmass}) at $\Delta m\to 0$) to form a singlet
($A_{\mu}^8$) and a triplet ($A_{\mu}^\alpha$) of $SU(2)_{C+F}$,
i.e. the spectrum of the theory becomes really non-Abelian in this
limit. Away from the Coulomb branch (at non-zero $\mu$) the presence
of this symmetry leads to emergence of the orientational zero modes
of the $\mathbb{Z}_2$-strings in the model, and the strings become
non-Abelian
\cite{HT1,ABEKY,HT2,SYmon}, see also the reviews
\cite{Trev,Jrev,SYrev,Trev2}. In what follows we consider the theory
with large number of flavors ($N_f=4,5$), and study whether the
theory at $r=2$ vacuum preserves its non-Abelian nature at $\Delta
m= 0$ as we reduce $m$ and go into strong coupling regime through
the monodromies which turn quarks into dyons.

In the case of coinciding masses the Higgs branches of the moduli
space can be also developed. The dimension of these
Higgs branches can be easily counted, see e.g. \cite{APS,MY}. For
example, from the root (\ref{avevna}) on the Coulomb branch one gets the
Higgs branch of dimension
\be
 {\rm dim} {\cal H}
=4N (N_f-N)\Big|_{N=2}=8(N_f-2)
\label{dimH}
\ee
which simply counts the difference between (real) number of
 scalar quarks and the number of F-term, D-term and gauge
constraints. The global unbroken group in this case is
$SU(2)_{C+F}\times SU(N_f-2)\times U(1)$.

In the case of large number of quarks $N_f \geq N_c$ the baryonic
branch can also appear, where the VEV's of the baryonic operators
\be
B \sim \epsilon^{A_1\ldots A_{N_c}}\epsilon_{k_1\ldots
k_{N_c}}Q^{k_1}_{A_1}\ldots Q^{k_{N_c}}_{A_{N_c}},
\ \ \  {\rm and}\ \ \
{\tilde B}
\sim \epsilon_{A_1\ldots A_{N_c}}\epsilon^{k_1\ldots
k_{N_c}}{\tilde Q}_{k_1}^{A_1}\ldots {\tilde Q}_{k_{N_c}}^{A_{N_c}}
\ee
do not vanish, in addition to the meson fields $M_A^B \sim {\tilde
Q}^B_k Q^k_A$. It happens, for example in the $SU(3)$ case (see
\rf{fir2}), if the flavor masses satisfy the constraint
\be
m_1+m_2+m_A = 0,\ \ \ A \neq 1,2
\label{baco}
\ee
and the third components $Q^3_A$ and ${\tilde Q}_3^A$ condense
\cite{APS,MY}. This phenomenon can happen, say, in the theory with $N_f=5$
flavors, with the pairwise coincident masses $m_1=m_3$ and $m_2=m_4$
but different $m_5$. However, we shall see in what follows, that the
baryonic branch constraint \rf{baco} does not intersect with the
strong-coupled domain in the mass plane and, therefore, does not
influence  our conclusions about the effective theory at strong
coupling.

\subsection{Non-Abelian semilocal strings}

Now we will briefly review some aspects of the non-Abelian strings
in our theory (\ref{qed}). The non-Abelian strings in
\ntwo QCD with $N_f=N$ arise, since the Abelian $\mathbb{Z}_N$-vortex solutions break
the $SU(N)_{C+F}$ global group, and therefore acquire the
orientational zero modes, associated with rotations of their color
flux inside the non-Abelian group $SU(N)$. The global group on the
$\mathbb{Z}_N$-string solution is broken  down to $SU(N-1)\times
U(1)$, and as a result, the moduli space of the non-Abelian string
is described by the coset space
\be
\frac{SU(N)}{SU(N-1)\times U(1)}\sim {\rm C}\mathbb{P}^{N-1}
\label{modulispace}
\ee
and the low-energy effective world sheet theory contains the
\ntwo SUSY two dimensional ${\rm C}\mathbb{P}^{N-1}$ sigma-model
\cite{HT1,ABEKY,SYmon,HT2}, see also \cite{MMY} about the
non-supersymmetric case.

If one adds the ``extra" quark flavors with degenerate masses, the
strings emerging in the theory with $N_f>N$ become semilocal, (see
\cite{AchVas} for a comprehensive survey of the Abelian
semilocal strings). It means, that the transverse size of such
string is no longer fixed, but becomes an additional modulus of a
string solution. In particular, the string solutions on the Higgs
branches (typical in the multiflavor theories) are commonly not
fixed-radius, but rather semilocal strings.

The non-Abelian semilocal strings in \ntwo QCD with $N_f>N$ (studied
in \cite{HT1,HT2,SYsem,Jsem}) have both orientational and size
moduli.
The effective two-dimensional theory
which describes the internal dynamics of the non-Abelian semilocal
string is \ntwot ``toric" sigma model, which includes fields
associated with both orientational and size zero modes of the string.
Its bosonic action in the gauge formulation (which
assumes taking the limit $e^2\to\infty$) has the form
\be
S = \int d^2 x \Big(
 \left|\nabla_i n^{P}\right|^2
 +\left|\tilde{\nabla}_i \rho^K\right|^2
 +\frac1{4e^2}(F_{ij})^2 + \frac1{e^2}\,
\left|\d_i\sigma\right|^2+ \frac{e^2}{2} \left(|n^{P}|^2-|\rho^K|^2 -2\beta\right)^2+
\\
+\sum_{P=1}^N\left|\sqrt{2}\sigma+m_P\right|^2
\left|n^{P}\right|^2 + \sum_{K=N+1}^{N_f}
\left|\sqrt{2}\sigma+m_K\right|^2\left|\rho^K\right|^2
\Big)
\label{wcp}
\ee
The world-sheet fields
$n^{P}$, $P=1,...,N$, and $\rho^K$, $K=N+1,...,N_f$ correspond to the orientational and
size moduli correspondingly, and have charges $+1$ and $-1$ with respect to the
auxiliary two-dimensional $U(1)$ gauge group, so that the covariant
derivatives, are $\nabla_i=\d_i-iA_i$ and
$\tilde{\nabla}_j=\d_j+iA_j$ respectively, where $i,j=1,2$.
 Small mass
differences $\left| m_A-m_B\right|$ lift orientational and size zero
modes generating a shallow potential on the moduli space.
The $D$-term condition
\be
  |n^P|^2 - |\rho^K|^2=2\beta\,,
\label{unitvec}
\ee
is implemented in the limit $e^2\to\infty$. Moreover, in this limit
the two-dimensional gauge field $A_i$  and its \ntwo bosonic
superpartner $\sigma$ become auxiliary and can be integrated out.
The two-dimensional FI D-term $\beta$ is related to the
four-dimensional coupling as
\be
\beta= \frac{2\pi}{g_2^2}
\label{betag}
\ee
We should mention here, that already the semilocality of strings destroys the confinement of monopoles
\cite{EY,SYsem}. The reason is that the transverse size of the string can grow indefinitely.
When it becomes comparable with the distance between the sources of magnetic flux (monopoles),
the linear confining potential between these sources is replaced by the Coulomb-like
potential, which does not confine. In order to preserve the confinement of monopoles by strings in our theory we should
lift the size zero modes keeping mass differences $|m_P-m_K|$, $P=1,...,N$, $K=N+1,...,N_f$
small, but nonvanishing.

The detailed discussion of the non-Abelian strings can be
found in \cite{Trev,Jrev,SYrev,Trev2}.
Here we would like just to point out, that they arise
in effective low-energy theories with actions like
\rf{u2model} basically independent of the nature of the fields.
We shall use this fact below, when discussing the effective gauge
theory of light dyons in the strong coupled domain for the original
theory \rf{u2model}. Much in the same way as for \rf{u2model},
the similar action of the effective theory has non-Abelian string
solutions, and their fluxes are determined by asymptotic of the gauge
fields at spatial infinity. The main question to be addressed below
is what are exactly the states, being confined by non-Abelian
semilocal strings in the dual low-energy theory.

\section{Semiclassical analysis of the mass plane
\label{ss:semi}}
\setcounter{equation}{0}

Our main interest in this paper is to study the basic features of
the non-Abelian confinement in Seiberg-Witten theory. It has been
proposed long ago \cite{MY}, that one way to do this is to
start with the $SU(N_c)$, $N_c\geq 3$ (at least $SU(3)$) gauge
theory in the ultraviolet, in order to be able to restore partially
the non-Abelian gauge group at the intermediate mass scale
$\sqrt{\mu m}\ll m$ in a weakly-coupled effective theory around some
${\cal N}=1$ vacuum. In order to get non-Abelian confinement one
should require sufficiently large number of flavors: it is
restricted from the top by common $N_f \leq 2N_c$, or $\beta_{\rm
UV} = 2N_c-N_f >0$ to have well-defined asymptotically free theory
in UV, but the effective theory should have $\beta_{\rm eff} =
\beta_{SU(2)} \leq 0$ in order to stay at weak coupling
\cite{APS} and to be able to trust
semiclassical string solutions, which ensure confinement. In
particular, for the $SU(3)$ supersymmetric QCD, broken down to the
only possible non-Abelian group $SU(2)$ in this case, one can
consider theory with $N_f=4$ or $N_f=5$ flavors. For further
simplicity, we shall take pairwise coinciding masses of flavor
multiplets, say $m_1=m_3$ and $m_2=m_4$, and therefore our ``phase
diagram'' would depend on only two parameters. The extra flavor mass
$m_5$ (in the $N_f=5$ theory; in $N_f=4$ theory it is absent)
 almost does not enter the game in semiclassical
regime: it is large and basically only renormalizes the scale
$\Lambda^2\equiv\left.\Lambda^2\right|_{N_f=4}\to
m_5\left.\Lambda\right|_{N_f=5}$. The difference of the exact
$N_f=5$ picture from that of $N_f=4$ we discuss below in
sect.~\ref{ss:nf5}.

\begin{figure}[tp]
\epsfysize=6cm
\centerline{\epsfbox{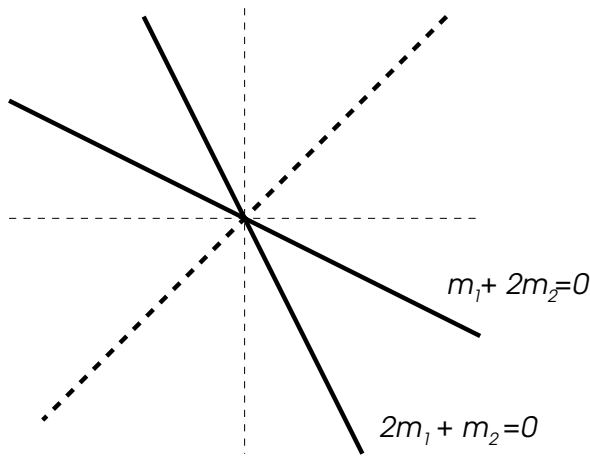}}
\caption{Semiclassical regime can reached at large values of masses and far from
two critical lines in the $(m_1,m_2)$-plane, depicted at this
picture.}
\label{fi:lines}
\end{figure}

Hence, one can start with large $m_{1,2} \gg \Lambda$ and stay at
quark $r=2$ vacuum, where the matrix of complex scalar from \ntwo
vector multiplet is taken in the form, say, \rf{fir2} in the
12-vacuum with the condensed two first flavors. Then $Q_A^k
\sim \delta_A^k \neq 0$, for $k,A=1,2$, solve the equations
$(\Phi^k_l+m_A\delta^k_l)Q_A^l = 0$, for the critical values of the
superpotential, i.e. the first and second massless quarks have the
color charges $\bmu_1$ and $\bmu_2$ from the triangle of fundamental
representation ${\bf 3}$ ($u$ and $d$ quarks in the terminology of
\cite{MY}) for the scalar matrix $\Phi$ is taken in the gauge
\rf{fir2} (we forget for a second about extra flavors with $A=3,4$,
and return to their influence to our conclusions later). The quark
condensate breaks gauge group completely on the scale $\langle Q_A
\rangle \sim \sqrt{\mu m_A}$, but at the intermediate energy scales
$\sqrt{\mu m} < E < m$ (when $m_1
\sim m_2 \sim m$) we get generically an effective $U(1)\times U(1)$
theory, but at $\Delta m =m_1-m_2\to 0$ this gauge group  enlarges
to $SU(2)\times U(1)\simeq U(2)$. The ``phase diagram'' in the
$(m_1,m_2)$-plane can be depicted as on fig.~\ref{fi:lines},
certainly this is only a ``real section'' of the two-dimensional
space of {\em complex} masses $(m_1,m_2) \in
\mathbb{C}^2$.

Three lines, depicted at fig.~\ref{fi:lines}, correspond to
coinciding eigenvalues of the matrix \rf{fir2}. The dashed line when
$m_1=m_2$ is the line of $SU(2)\times U(1)$ theory with $N_f=4$
light fundamental multiplets, charged w.r.t. $SU(2)$ group.
The effective action of this theory is given by eq.~(\ref{u2model}). It has
therefore $\beta_{\rm eff} = \beta_{SU(2)} = 4-N_f =0$, and stays at
weak coupling, if we fix non-running coupling to be small ${1\over
g^2}\sim\log{m\over\Lambda}\gg 1$ at the scale, when $SU(3)$ breaks
down to $SU(2)\times U(1)$. The corresponding $SU(2)$ subgroup
corresponds to the $\balpha_{12}$-direction on fig.~\ref{fi:su3},
and we have doublet of light quarks ($\bmu_1$ and $\bmu_2$) in the
fundamental representation of this $SU(2)$.

\begin{figure}[tp]
\epsfysize=8cm
\centerline{\epsfbox{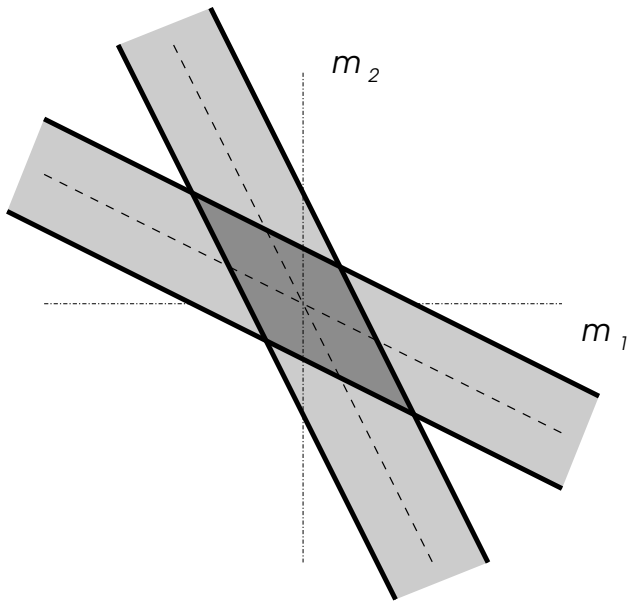}}
\caption{Fine structure of the (real slice of) the picture at fig.~\ref{fi:lines} in
$(m_1,m_2)$-plane, when the equations of border lines \rf{mlines}
are replaced by $m_1+2m_2=\pm 2\Lambda$ and $2m_1+m_2=\pm 2\Lambda$.
Light regions correspond to weak coupling regime with condensates of
two light quarks. In light-grey regions one deals with condensed
(mutually orthogonal) light quarks and dyons, while in the dark
region - rhombus around the origin, we get a doublet of light
dyons.}
\label{fi:dia_yu}
\end{figure}

Two solid lines
\be
\label{mlines}
2m_1+m_2=0,\ \ \ m_1+2m_2=0
\ee
are different: here, in each case, one deals with restoration of a
$SU(2)$ subgroup, now either in $\balpha_1$ or in $\balpha_2$
direction (fig.~\ref{fi:su3}). Each ($\balpha_1$- or $\balpha_2$-)
$SU(2)$ subgroup interacts with only two charged light flavors, with the
charges given either by $\bmu_1$ or by $\bmu_2$, due to
$\bmu_i\cdot\balpha_j=\delta_{ij}$ (we neglect the contribution of
the heavy quark fields with masses of the order of $m \gg
\Lambda$). Hence, the corresponding effective beta-function
$\beta_{\rm eff} = \beta_{SU(2)} = 4-N_f =2
>0$, and the effective theory falls into strong-coupling regime by
the Seiberg-Witten mechanism. Thus, our semiclassical considerations
are not reliable in the vicinity of \rf{mlines}, and we will have to
extract more information from the exact solutions to study the
vicinity of these lines: the ``fat lines'' with the width of the
order of $\Lambda$, see fig.~\ref{fi:dia_yu}.

Remember now, that the picture at fig.~\ref{fi:lines} is in fact
real slice of the full complex picture, and each straight line has
real codimension 2. It means, that one can go around each solid
line, naively separating two ``weakly coupled'' sectors, moving to
complex domain and keeping $|m_A| \gg \Lambda$, $A=1,2$, taking into
account the weak-coupling monodromies. Restricting ourselves to sit
in the $r=2$ vacuum (where first two flavors condense) in one of the
sectors, we shall preserve this condition in all weakly coupled
domains, once it had been chosen (12-vacuum for our choice). The
consistency of this picture is better seen in different from
\rf{fir2} gauge, when the (real) eigenvalues of $\Phi$ are ordered
(see Appendix~\ref{app:gauge}), but we shall use mostly the gauge
\rf{fir2} below, as more adequate for our purposes.

As we already noticed, at the vicinity of solid lines from
fig.~\ref{fi:lines} the theory around $r=2$ vacuum falls into the
strong coupling regime. Differently one can say, that on these lines
$r=2$ vacuum collides with on of the $r=1$ vacua, since matrix
\rf{fir2} acquires simultaneously the form
\be
\label{fir1}
\left.\Phi\right|_{2m_1+m_2=0}
= -\left(
\begin{array}{ccc}
  -{m_2\over 2} &  &  \\
   & m_2 &  \\
   &  & -{m_2\over 2}
\end{array}\right),\ \ \ \
\left.\Phi\right|_{m_1+2m_2=0}
= -\left(
\begin{array}{ccc}
    m_1 &  & \\
   &  -{m_1\over 2} &  \\
   &  & -{m_1\over 2}
\end{array}\right)
\ee
The theory at $r=1$ vacuum has, in addition to the light quark, a
light monopole or dyon, with the electric and magnetic charges
$(n_e,n_m)=(0,1)$ and $(n_e,n_m)=(1,1)$ in units of the orthogonal root to the charge of
quark (see detailed discussion in \cite{MY}). The semiclassically
unbroken $SU(2)$-subgroups in \rf{fir1} are ``broken back'' to
$U(1)$ on the scale $\Lambda$, say
\be
\label{fir1l}
\left.\Phi\right|_{r=1}
\sim\left(
\begin{array}{ccc}
  -{m_2\over 2}\pm\Lambda &  &  \\
   & m_2 &  \\
   &  & -{m_2\over 2}\mp\Lambda
\end{array}\right),
\ee
 i.e. the $SU(2)\subset SU(3)$ subgroup exists
only for $\Lambda\to 0$, while the $\Lambda$-splitting in \rf{fir1}
is governed at strong coupling by the original Seiberg-Witten
mechanism \cite{SW1}. Therefore, the exact behavior around the solid
lines at fig.~\ref{fi:lines} is describe by the effective
strongly-coupled $SU(2)$ gauge theory with $N_f=2$ light fundamental
multiplets, see Appendix~\ref{app:deg} and sect.~\ref{ss:pbp}.

This transparent example obeys, however, an important internal
problem. When the UV gauge group $SU(3)$ breaks at the scale $m$
into $SU(2)\times U(1)$, the $SU(2)$ theory with $N_f=4$ flavors is
conformal, and its coupling does not run, being fixed by
\be
e^{-{8\pi^2\over g^2}} = \left({\Lambda\over m}\right)^{2N_c-N_f} =
\left({\Lambda\over m}\right)^2
\label{su2c}
\ee
If average mass $m$ goes below $\Lambda=\Lambda_{SU(3)}$, the
effective low-energy conformal $SU(2)$ theory we consider is in the
strong coupling regime, so that the naive semiclassical analysis is
not applicable. The coupling of quantum $SU(2)$ theory with $N_f=4$
is being moreover renormalized by the instanton effects (see e.g.
\cite{Khrev,KlMa,Zam_agt}), and therefore goes beyond our control at
$m\leq \Lambda$.

In order to avoid this one has to consider instead the theory with
$N_f=5$, which becomes the IR free in the $SU(2)$ sector. In this
theory we shall always take $m_A \sim m$, and $|\delta m_{AB}| \ll
|m|$, for any $A,B=1,\ldots,N_f=5$. At energies $E>|m|$ this is an
asymptotically free $SU(3)$ \ntwo supersymmetric gauge theory with
$\beta_3 = 6-N_f=1$, i.e. $1/g_3^2\sim (\log E/\Lambda)$, while at
$E<|m|$ this is a zero-charge $SU(2)$ theory with $\beta_2 =
4-N_f=-1$, or $1/g_2^2\sim -\log (E/{\tilde\Lambda})$. At $E=m$
these two lines intersect, what gives
\be
\label{tlam}
{\tilde\Lambda} = {\rm max}\left({m^2\over\Lambda},\Lambda\right),
\ee
where we have also taken into account that at $|m|<\Lambda$ the scale of
breaking of the $SU(3)$ gauge symmetry down to $U(2)$ is determined by $\Lambda$.

Below we  consider the case, when $m_1=m_3$, and $m_2=m_4$,
while $m_5$ can vary in different ranges. More precisely
our low energy $U(2)$ theory with five flavors is not
asymptotically free and stays at weak coupling if
\be
|m_A-m_B|\ll \tilde{\Lambda},\ \ \ \  A,B=1,2,5
\label{semior}
\ee
i.e. all mass differences are essentially small.

\section{Exact solution of $N_c=3$ and $N_f=4,5$ theories
\label{ss:exact}}
\setcounter{equation}{0}

\subsection{Seiberg-Witten theory for supersymmetric QCD
\label{ss:swfund}}

Generic curve for \ntwo supersymmetric QCD with $N_c$ colors and
$N_f$ flavors can be written in the form
\cite{HO,ArFa,KLTY,ArPlSh}
\be
\label{cuy}
y^2=P(x)^2-4Q(x)
\ee
where
\be
\label{PQ}
P(x)=\prod_{i=1}^{N_c}(x-\phi_i), \ \ \
\sum_{i=1}^{N_c}\phi_i=-\Lambda\delta_{N_f,2N_c-1}
\\
Q(x)=\Lambda^{2N_c-N_f}\prod_{A=1}^{N_f}(x+m_A)
\ee
with two polynomials of powers $N_c$ and $N_f$ respectively.
Semiclassically the roots $\{ \phi_i \}$, $i=1,\ldots,N_c$ coincide
with the eigenvalues of the matrix $\Phi$ of the condensate of the
complex scalar from the vector multiplet of \ntwo supersymmetric
Yang-Mills theory, but being computed exactly they (or, better,
their symmetric functions) are got corrected in (dependent upon
$\Lambda$ and $m_A$ way) due to the instanton effects. The curve
\rf{cuy} can be also re-written as \cite{GMMM}
\be
\label{cuw}
w+{Q(x)\over w}=P(x)
\ee
or
\be
\label{cuW}
W+{1\over W}={P(x)\over\sqrt{Q(x)}}
\ee
with $y=w-{Q(x)\over w} = \left(W-{1\over W}\right)\sqrt{Q(x)}$.

The curves \rf{cuy}, \rf{cuw} or \rf{cuW} are endowed with a
generating differential
\be
\label{dS}
dS \sim x{dw\over w} = x{dW\over W}+ \frac12 x{dQ\over Q} = {x
dP\over y} - x {P\over 2y}{dQ\over Q}+ \frac12 x{dQ\over Q}
\ee
where it is chosen to have the residues
\be
\label{resdS}
\res_{P_A^\pm}dS =  m_A \cdot \left.{P\over 2y}\right|_{x=-m_A}-{m_A\over 2} =
-m_A
\ee
at the points $P_A
$ with $x(P_A^\pm)=-m_A$ at one of the sheets of \rf{cuy}. The
variation of
\rf{dS} at constant $W$ gives rise to
\be
\label{vardS}
\delta dS \sim {dx\over y}\left(\delta P(x) -
\half P{\delta Q(x)\over Q(x)}\right)
\ee
In the case of $SU(3)$ gauge group it is convenient to introduce
explicitly
\be
P(x)=(x-\phi_1)(x-\phi_2)(x-\phi_3) = x^3-ux-v
\label{pho}
\ee
with, for $N_f<5$ and $\phi_3=-\phi_1-\phi_2$,
\be
\label{uvphi}
u=\phi_1^2+\phi_2^2+\phi_1\phi_2
\\
v=-\phi_1\phi_2(\phi_1+\phi_2)
\ee
so that for the Seiberg-Witten periods one gets
\be
\label{SWper}
a_i={1\over 2\pi i}\oint_{A_i}dS,\ \ \ \ a^D_i={1\over 2\pi
i}\oint_{B_i}dS
\ee

\subsection{Conformal $N_f=4$ theory}

First we consider the $N_f=4$ case with the pairwise coinciding
masses, when the Seiberg-Witten curve becomes
\be
\label{su3nf4}
y^2= (x-\phi_1)^2(x-\phi_2)^2(x+\phi_1+\phi_2)^2-4\Lambda^2(x+m_1)^2(x+m_2)^2,
\ee
To describe the 12-vacuum exactly in terms of (\ref{su3nf4}) it is
necessary to ensure that this curve has two double roots, determined
by the quark masses in semiclassical limit. This is easily obtained
- for the pairwise coinciding masses - by putting {\em exactly} any
two roots of the polynomial
\rf{pho} to coincide with the masses $m_{1,2}$ with the opposite
sign. For  \rf{fir2}, this is
\be
\phi_1 = -m_1,\ \ \ \ \ \phi_2 = -m_2
\label{firstphi}
\ee
so that the curve \rf{su3nf4} turns into
\be
\label{nf4deg}
y^2 = (x+m_1)^2(x+m_2)^2\left((x-m_1-m_2)^2-4\Lambda^2\right) =
\\
= (x+m_1)^2(x+m_2)^2\left((x-M)^2-4\Lambda^2\right)\equiv
(x+m_1)^2(x+m_2)^2Y^2
\\
M=m_1+m_2
\ee
where
\be
Y^2=(x-M)^2-4\Lambda^2
\ee
coincides with the formal "pure $U(1)$" \ntwo SUSY gauge theory
curve \cite{LMN,MN}, with the only VEV given here by $M=m_1+m_2$.

\begin{figure}[tp]
\epsfysize=6.5cm
\centerline{\epsfbox{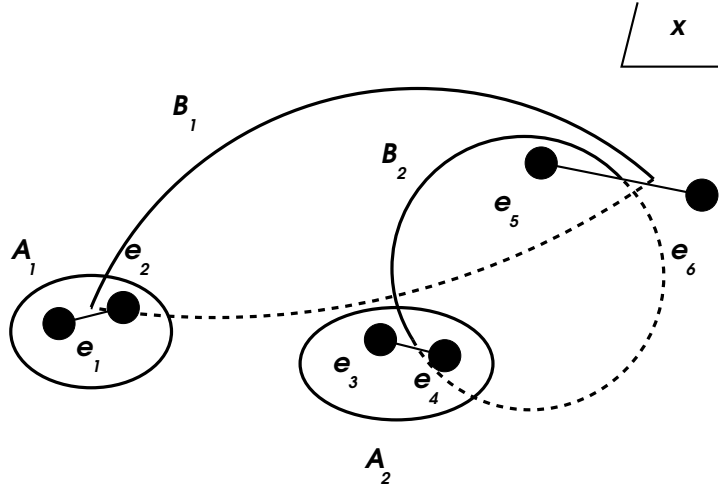}}
\caption{Homology basis for the $SU(3)$ curve in complex $x$-plane.}
\label{fi:hom}
\end{figure}
Two contours $A_1$ and $A_2$ shrink for the curve \rf{nf4deg}, (see
fig.~\ref{fi:hom}), and the associated periods $a_{1,2}={1\over 2\pi
i}\oint_{A_{1,2}} dS$ (and therefore $a_3$ and $a_8$, see
\rf{aw}) reduce to the residue integrals ( see
Appendix~\ref{app:deg}). Here we choose canonical
basis of the $(A,B)$-cycles on
\rf{su3nf4} as follows: the $A_k$-cycle surrounds the cut, which
shrinks to $x=\phi_k\simeq -m_k$, both for $k=1,2$, while the dual
$B_k$-cycles obey $A_i \circ B_j =\delta_{ij}$, see
fig.~\ref{fi:hom}. The values of the residues of the differential
\rf{dS} on the curve \rf{nf4deg}  correspond to exact vanishing of the effective masses
of the quarks $A=1,2$ in the 12-vacuum (at $\mu=0$, see
\rf{aw})
\be
 a_1 + m_1 = {a_3\over\sqrt{2}}+ \frac{a_8}{\sqrt{6}} +m_1=0
\\
 a_2 + m_2 = -{a_3\over\sqrt{2}}+ \frac{a_8}{\sqrt{6}} +m_2=0
\label{mlq}
\ee
In \rf{mlq} we have used that the charges of these two quarks in the
gauge \rf{fir2} are given by two weights of the algebra (see
fig.~\ref{fi:su3})
\be
Q^1_1:\ \ \ {\bf n}_e = {\bmu_1\over\sqrt{2}},\ {\bf n}_m=0\ \ \ \
{\rm or} \ \ \ \ \left(n_e^3,n_m^3;\,n_e^8,n_m^8\right)=
\left(\frac12,0;\,\frac1{2\sqrt{3}},0\right)
\label{qc11}
\ee
and
\be
Q^2_2:\ \ \ {\bf n}_e = {\bmu_2\over\sqrt{2}},\ {\bf n}_m=0\ \ \ \
{\rm or} \ \ \ \
\left(n_e^3,n_m^3;\,n_e^8,n_m^8\right)=
\left(-\frac12,0;\,\frac1{2\sqrt{3}},0\right)
\label{qc22}
\ee
of the $SU(3)$ gauge group, broken down generally to $U(1)\times
U(1)$ by nonvanishing $\Delta m_{AB}\neq 0$. Here $n_e^3,n_m^3$ and $n_e^8,n_m^8$
are electric and magnetic charges of a state with respect to two Cartan generators of $SU(3)$
gauge group, $T^3$ and $T^8$ respectively.

The main outcome of this solution is absence of any corrections of
the order of $\Lambda$ to the first two $\phi$'s in
(\ref{firstphi}), which means that in the equal mass limit these two
$\phi$'s become exactly equal. This is a signal of restoration of
the non-Abelian $U(2)\simeq SU(2)\times U(1)$ gauge group at the
root of the Higgs branch (at $\mu=0$). We have expected this in
sect.~\ref{ss:semi} in semiclassical at large masses $m$. Now we
see, that this phenomenon occurs for arbitrary $m$, in particular,
if we reduce $m$ and go all the way to the strong coupling
region\footnote{This is in perfect agreement with
\cite{APS}, where non-Abelian dual gauge groups $U(r)$ were
identified at the roots of non-baryonic Higgs branches in the
$SU(N)$ gauge theory with $N_f$ massless quarks, $r<2N_f$.} at
$m\ll\Lambda$. As was already mentioned, the physical reason for the
emergence of the non-Abelian  gauge group is that the dual
low-energy effective theory with the dual gauge group $U(2)\simeq
SU(2)\times U(1)$ at $m\ll\Lambda$ is not asymptotically free in the
equal mass limit and stays at weak coupling\footnote{Strictly
speaking, this is true only for the theory with $N_f=5$, which we
consider in sect.\ref{ss:nf5}.}. Therefore, the classical analysis
showing that the non-Abelian gauge group is restored at the root of
the Higgs branch remains intact in quantum theory.

The curve \rf{nf4deg} has two double roots at
\be
e_1 = e_2=-m_1,\ \ \ \ \ e_3 = e_4=-m_2.
\label{roots14}
\ee
while the remaining two roots of the "$U(1)$-curve" are at
\be
e_5  = m_1+m_2 - 2\Lambda,\ \ \ \ \ e_6=m_1+m_2 + 2\Lambda,
\label{roots56}
\ee
In the monopole singularity the other roots coincide, e.g.
$e_2=e_5$, see fig.~\ref{fi:hom}, and the $B_1$-contour shrinks
producing a regular period. Indeed (cf. with \rf{aDm}),
\be
\label{B1dg}
a_1^D = {a^D_3\over\sqrt{2}}+\sqrt{\frac{3}{2}}a^D_8 = {1\over 2\pi
i}\oint_{B_1}dS =
\\
= -{i\over \pi}\left(\sqrt{(2m_1+m_2)^2-4\Lambda^2}
+\left(m_1+{m_2\over
2}\right)\log{2m_1+m_2-\sqrt{(2m_1+m_2)^2-4\Lambda^2}\over
2m_1+m_2+\sqrt{(2m_1+m_2)^2-4\Lambda^2}}\right)
\ee
Again, we fix its real part to ensure that \rf{B1dg} vanishes at
$2m_1+m_2=2\Lambda$, corresponding to the masslessness of the
monopole with the charges
\be
{\bf n}_e = 0,\ {\bf n}_m={\balpha_1\over\sqrt{2}}\ \ \ \ {\rm or} \
\ \ \
\left(n_e^3,n_m^3;\,n_e^8,n_m^8\right)=
\left(0,\frac12;\,0,\frac{\sqrt{3}}{2}\right).
\label{1mnc}
\ee
which is one of three $SU(3)$ elementary monopoles, whose charges
are determined by the roots of the $su(3)$ algebra, see
fig.~\ref{fi:su3}. Exchanging in \rf{B1dg} $m_1\leftrightarrow m_2$,
one gets at $m_1+2m_2=2\Lambda$ vanishing of the mass
\be
a^D_2=-{a^D_3\over\sqrt{2}}+\sqrt{\frac{3}{2}}a^D_8= {1\over 2\pi
i}\oint_{B_2}dS =
\\
= -{i\over \pi}\left(\sqrt{(m_1+2m_2)^2-4\Lambda^2} +\left({m_1\over
2}+m_2\right)\log{m_1+2m_2-\sqrt{(m_1+2m_2)^2-4\Lambda^2}\over
m_1+2m_2+\sqrt{(m_1+2m_2)^2-4\Lambda^2}}\right)
\label{B2dg}
\ee
of the monopole with the charge
\be
{\bf n}_e = 0,\ {\bf n}_m={\balpha_2\over\sqrt{2}}\ \ \ \ {\rm or}
\ \ \ \
\left(n_e^3,n_m^3;\,n_e^8,n_m^8\right)=
\left(0,-\frac12;\,0,\frac{\sqrt{3}}{2}\right).
\label{2mnc}
\ee
Clearly, the imaginary part $\Im(a_1^D) =0$ vanishes also at
$2m_1+m_2=-2\Lambda$, and similarly $\Im(a_2^D) =0$ if
$m_1+2m_2=-2\Lambda$. However, the real parts of expression
\rf{B1dg} at $2m_1+m_2=-2\Lambda$ equals to $2m_1+m_2 = -
\balpha_1\cdot{\bf a}$, or to the mass of the W-boson with the charge $\balpha_1$.
Hence, at $2m_1+m_2=-2\Lambda$ one gets the massless dyon with the
charge
\be
{\bf n}_e = {\balpha_1\over\sqrt{2}},\ {\bf
n}_m={\balpha_1\over\sqrt{2}}\
\ \ \ {\rm or}
\ \ \ \
\left(n_e^3,n_m^3;\,n_e^8,n_m^8\right)=
\left(\frac12,\frac12;\,\frac{\sqrt{3}}{2},\frac{\sqrt{3}}{2}\right).
\label{a1dyc}
\ee
and similarly, the massless dyon with the charge
\be
{\bf n}_e = {\balpha_2\over\sqrt{2}},\ {\bf
n}_m={\balpha_2\over\sqrt{2}}\
\ \ \ {\rm or}
\ \ \ \
\left(n_e^3,n_m^3;\,n_e^8,n_m^8\right)=
\left(-\frac12,-\frac12;\,\frac{\sqrt{3}}{2},\frac{\sqrt{3}}{2}\right).
\label{a2dyc}
\ee
at $m_1+2m_2=-2\Lambda$.

\subsection{Permutation of the branch points
\label{ss:pbp}}

Vanishing of the monopole masses $a^D_1$ by \rf{1mnc} at
$2m_1+m_2=2\Lambda$ and, similarly, of $a^D_2$ at
$m_1+2m_2=2\Lambda$ leads to appearance of massless monopoles in
12-vacuum with massless quarks \rf{mlq}, what means that we arrive
at the Argyres-Douglas (AD) points/lines
\cite{AD,APSW}. The AD point corresponds to particular value of the mass
parameters (or a curve in the space of all mass parameters, as in
our theory), where the mutually nonlocal states become massless
simultaneously. From (\ref{roots14}) and (\ref{roots56}) one finds,
that in our $N_c=3$, $N_f=4$ theory there are four AD lines where
the $r=2$ 12-vacuum collides with the magnetic singularities. These
are at
\be
\label{AD1}
2m_1+m_2= 2\Lambda,\quad e_1=e_2=e_5=-m_1
\\
2m_1+m_2= -2\Lambda,\quad e_1=e_2=e_6=-m_1
\ee
and at
\be
\label{AD2}
2m_2+m_1= 2\Lambda,\quad e_3=e_4=e_5=-m_2
\\
2m_2+m_1= 2\Lambda,\quad e_3=e_4=e_6=-m_2
\ee
producing the quantum regularization (depicted at
fig.~\ref{fi:dia_yu}) of two classical lines \rf{mlines} from
fig.~\ref{fi:lines} where the semiclassical approximation is no
longer valid.

As we reduce the masses $m$ and pass from weak coupling into the
strong coupling domain, crossing \rf{AD1} and/or \rf{AD2}, along the
Coulomb branch at $\mu=0$, the quantum numbers of massless quarks
$Q^1_1$ and $Q^2_2$ change due to nontrivial monodromies in the
space of masses\footnote{ Note that $U(2)$ theory with $N_f=4,5$
does not have monodromies, associated with $\Delta m_{AB}$ since
this theory is not asymptotically free, to be compared with studied
in
\cite{SYcross,SYdual}.}. The complex $m$-plane has cuts, crossing
these cuts the periods $a$ and $a^D$ change linearly, accordingly
changing the quantum numbers of corresponding states (there
monodromies were studied in
\cite{BF} in the $SU(2)$ gauge theory through a
monodromy matrix approach). Let us demonstrate now, that as we pass
through the AD lines \rf{AD1}, \rf{AD2} the root pairings on the
curve \rf{su3nf4} change and light quarks transform into the light
dyons.

Take real $m_1$ and  $m_2$, and consider the first AD line in (\ref{AD1}).
On this line the 12-vacuum with two massless quarks
\rf{qc11} and \rf{qc22} collides with $r=1$ vacuum where quark $Q^2_2$ with
the charges \rf{qc22} and the monopole (\ref{1mnc}) are massless.
The roots $e_3$ and $e_4$ are far away and, therefore, the charges
of the $Q^2_2$ quark \rf{qc22} do not change, and we focus on the
colliding roots $e_1$, $e_2$ and  $e_5$, $e_6$ (see \rf{AD1}) as we
decreasing $m$. Hence, this situation is described by an effective
$SU(2)$ curve with four roots and two cuts, and it is described in
detail in Appendix~\ref{app:su2cu}.

In order to see this, let us slightly split the degenerate branch
points of
\rf{nf4deg} by shifting $\phi_1$ from its solution \rf{firstphi},
parameterizing the shift as
\be
\label{cpsh}
x = -m_1 +z,\ \ \ \phi_1 = -m_1 + \delta\phi,\ \ \ 2m_1+m_2=2\Lambda+\epsilon,
\ee
and relabel the roots \rf{roots14}, \rf{roots14} after the shift
$e_i\to z_i=e_i+m_1$, $i=1,\ldots,6$. The curve \rf{su3nf4} now
acquires the form
\be
\label{cu4}
y^2 = (m_1-m_2)^2\left[(z-\delta\phi)^2(z-\epsilon-2\Lambda+\delta\phi)^2-
4\Lambda^2z^2\right] \sim
\\
\sim \left[z^2 - (4\Lambda +\epsilon) z +2\Lambda \delta\phi\right]
\left[z^2 - z\epsilon+ 2\Lambda \delta\phi\right]
\ee
where we have omitted the overall factor and inessential terms one
can neglect in this approximation. The polynomial in the r.h.s. of
\rf{cu4} obviously factorizes into the product of two quadratic
polynomials, whose roots are the ends of the cut, surrounding
the $A_1$ and $A_3$ cycle (see fig.~\ref{fi:hom}), while the
degenerate $A_2$-cut is separated from
\rf{cu4} for a while, i.e. the situation is indeed
reduced to the curve of effective $SU(2)$ theory with two flavors,
see details in Appendix~\ref{app:su2cu}. One can study therefore
with the help of \rf{cu4} what happens literally, when the ends of
the first and third cuts touch each other.

Comparing the last relation from \rf{cpsh} with
\rf{AD1} one finds that there are two critical regimes for
parameter $\epsilon$ in \rf{cu4}, the first is at $\epsilon\sim 0$
or the vicinity of upper AD-line (the first equation in
(\ref{AD1})), while in the vicinity of lower AD-line (the second
equation in (\ref{AD1})) one should take $\epsilon = -4\Lambda +
{\tilde\epsilon}$, with ${\tilde\epsilon}\sim 0$.

Consider, first, $\epsilon\sim 0$ (and let us be interested in
domain, where $z \ll \Lambda$).
Equation \rf{cu4} turns into
\be
\label{cu4e1}
y^2
\sim \left(z^2 - 4\Lambda z+\delta\right)
\left(z^2 - \epsilon z+\delta\right)
\ee
with $\delta\equiv 2\Lambda\delta\phi$. It has "large root" $z_6
\approx 4\Lambda$ of the first quadratic polynomial, which is far
away from its second root $z=z_1\approx\half\delta\phi\approx 0$,
they correspond to $e_6$ and $e_1$ respectively in \rf{AD1}. Two roots of the
second quadratic polynomial at
\be
\label{z1pm}
z_{5,2} = {\epsilon\over 2} \pm \sqrt{{\epsilon^2\over 4}-\delta}
\ee
At $\epsilon>0$ we have $z_1\approx z_2$ ( for the minus sign in
\rf{z1pm}), while $z_5\approx\epsilon$. The contour $A_1$ goes
therefore around a "small" cut, and the corresponding period
integral is associated with the mass $a_1+m_1$ of light quark
$Q^1_1$, while the roots $z_6\approx 4\Lambda$ and $z_5 \approx
\epsilon$ are the ends of the cut, surrounded by $A_3$-cycle.

At $\epsilon\to 0$ the $B_1$-cycle also degenerates, which means
that monopole with the charge (\ref{1mnc}) also becomes massless.
Then, when $\epsilon<0$ becomes negative, the picture changes
drastically, now two roots at $z=z_1\approx 0$ and $z=z_5$ become
close to each other, while $z=z_2$ is negative of the order of
$|\epsilon|$. After such transition we get a small $A_1+B_1$-cycle,
with the mass of light state, corresponding to the period integral
along the $A_1+B_1$-contour, see Fig.~\ref{fi:hom} (see also details in
Appendix~~\ref{app:su2cu}).

This means, that the massless quark  $Q^1_1$ transforms into the
massless dyon $D^1_1$ with the quantum numbers
\be
D^1_1:\ \ \ {\bf n}_e = {\bmu_1\over\sqrt{2}},\ {\bf
n}_m={\balpha_1\over\sqrt{2}}\ \ \ \ {\rm or} \ \ \ \
\left(n_e^3,n_m^3;\,n_e^8,n_m^8\right)=
\left(\frac12,\frac12;\,\frac1{2\sqrt{3}},\frac{\sqrt{3}}{2}\right),
 \label{dyon1}
\ee
or ${\bf n}^1 = {1\over\sqrt{2}}(\bmu_1\oplus\balpha_1)$, while the
charge \rf{qc22} of the quark $Q^2_2$ does not change. Hence, the
quantum numbers of the massless quark $Q^1_1$  in the $r=2$ vacuum,
after the collision with the monopole singularity, become shifted by
the monopole magnetic charge.

Analogously, one can analyze the first AD line in (\ref{AD2}) where
the $r=2$ vacuum collides with another $r=1$ vacuum, containing
massless quark $Q^1_1$ and the monopole with the charge
\rf{2mnc}. Now $Q^1_1$
does not change its charge, while the quark $Q^2_2$ acquires
additionally the charge of the monopole (\ref{2mnc}). As a result,
below both AD lines, i.e. inside the rhombus in
fig.~\ref{fi:dia_yu}, we end up with two massless dyons: the dyon
$D^1_1$ with the charge ${\bf n}^1 =
{1\over\sqrt{2}}(\bmu_1\oplus\balpha_1)$
\rf{dyon1}, and
\be
  D^2_2:\ \ \
{\bf n}_e = {\bmu_2\over\sqrt{2}},\ {\bf
n}_m={\balpha_2\over\sqrt{2}}\
\ \ \ {\rm or} \ \ \ \
\left(n_e^3,n_m^3;\,n_e^8,n_m^8\right)=
\left(-\frac12,-\frac12;\,\frac1{2\sqrt{3}},\frac{\sqrt{3}}{2}\right),
\label{dyon2}
\ee
or ${\bf n}^2 = {1\over\sqrt{2}}(\bmu_2\oplus\balpha_2)$. The quark
masslessness conditions  (\ref{mlq}) at small $m$, inside the
rhombus, are replaced by the dyon masslessness conditions, namely,
\be
a_1+a^D_1+m_1 = \bmu_1\cdot{\bf a}+\balpha_1\cdot{\bf a}^D+m_1 =
{a_3\over\sqrt{2}}+ {a^D_3\over\sqrt{2}}+ {a_8\over\sqrt{6}}
+\sqrt{\frac{3}{2}} a^D_8 +m_1=0
\\
a_2+a^D_2+m_2 = \bmu_2\cdot{\bf a}+\balpha_2\cdot{\bf a}^D+m_2 = -
{a_3\over\sqrt{2}}- {a^D_3\over\sqrt{2}}+ {a_8\over\sqrt{6}}
+\sqrt{\frac{3}{2}} a^D_8 +m_2=0
\label{mld}
\ee
Let us point out here the crucial fact, that both electric and
magnetic charges of our massless dyons \rf{dyon1}, \rf{dyon2} are
$\pm \half$ with respect to the $\tau^3$-generator of the dual
$U(2)$ gauge group, i.e. they can belong to the fundamental
representation of this group; moreover, all dyons $D^l_A$ ($l=1,2$)
form the color doublets. This is another confirmation of the
conclusion we already made above, that the non-Abelian $SU(2)$
factor of the dual gauge group gets restored in the equal mass
limit.

Going further, and crossing the lower AD lines at
fig.~\ref{fi:dia_yu}, corresponding to the lower equations in
\rf{AD1}, \rf{AD2}, one comes back to the 12-vacuum we had in weak
coupling, where the electric $A$-cycles are small. The dyons
$D^k_A$, $k,A=1,2$ with the charges \rf{dyon1} and \rf{dyon2} change
their quantum numbers on these lines due to the presence of massless
dyons with the charges $\balpha_1\oplus\balpha_1$ and
$\balpha_2\oplus\balpha_2$ (on \rf{AD1} and \rf{AD2}
correspondingly), whose both electric and magnetic constituents are
given entirely in terms of the {\em root} vectors, and turn back
into the light quarks. The detailed analysis can be found in
Appendices~\ref{app:deg},\ref{app:su2cu}, and we skip it in the main
text.

\subsection{$N_f=5$ IR free theory
\label{ss:nf5}}

Let us turn now to the peculiarities of the above analysis in the
$N_f=5$ case. If $m_1=m_3$ and $m_2=m_4$ the $N_f=5$ curve
\rf{cuy}, \rf{PQ} acquires the form
\be
\label{su3nf5}
y^2=
(x-\phi_1)^2(x-\phi_2)^2(x+\phi_1+\phi_2+\Lambda)^2-4\Lambda(x+m_1)^2(x+m_2)^2(x+m_5)
\ee
and if we are exactly in the $r=2$ 12-vacuum, similarly to the
$N_f=4$ case it degenerates to
\be
\label{nf5dg}
y^2= (x+m_1)^2(x+m_2)^2Y^2
\ee
where now
\be
\label{u1n1}
Y^2 = p^2 - 4\Lambda(x+m_5)
\\
p = x-M+\Lambda = x-m_1-m_2+\Lambda
\ee
is the curve of the "formal $U(1)$" theory with one flavor.

\begin{figure}[tp]
\epsfysize=9cm
\centerline{\epsfbox{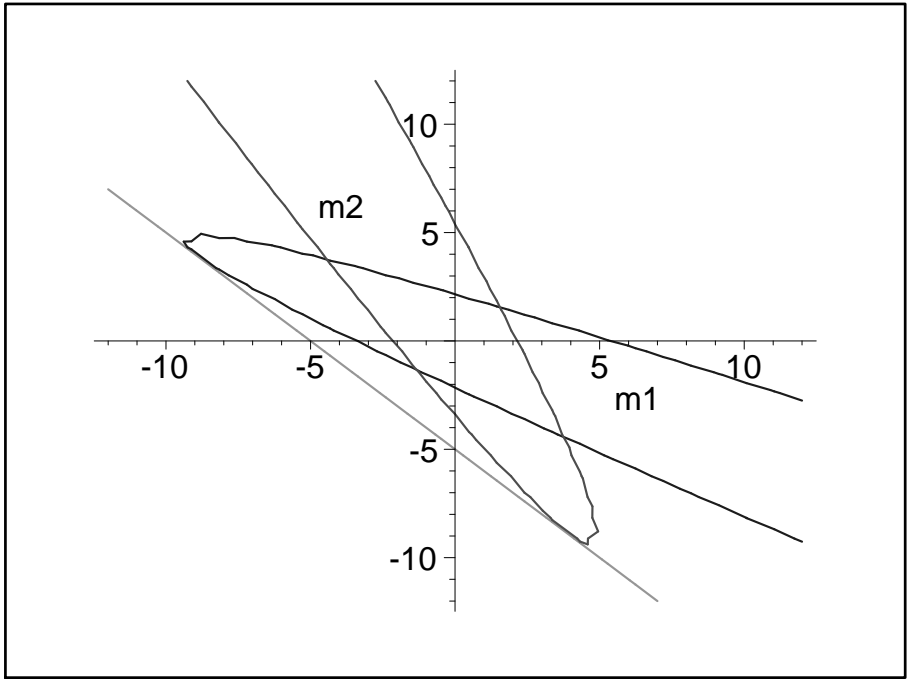}}
\caption{Weak-coupling parabolas in $N_f=5$ theory. Here
the curves \rf{par1} and \rf{par2} are plotted at $m_5=5\Lambda$,
and the straight line $\Sigma=m_1+m_2+m_5=0$ is the baryonic branch
condition, see \rf{Dnf5}.}
\label{fi:parw}
\end{figure}
Compare to the $N_f=4$ case, the roots \rf{roots14} for the curve
\rf{nf5dg} remain intact, while instead of \rf{roots56}, one gets
from \rf{u1n1}
\be
e_5=M+\Lambda-2\sqrt{\Lambda \Sigma},\ \ \
e_6=M+\Lambda+2\sqrt{\Lambda \Sigma}
\label{56roots}
\ee
with $M=m_1+m_2$ and
\be
\label{Dnf5}
\Sigma = M+m_5 = m_1+m_2+m_5
\ee
is the discriminant of \rf{u1n1}, whose vanishing means that we come
to the origin of the baryonic branch.

It is easy to see now, that instead of the AD straight lines
\rf{AD1} and \rf{AD2} of the $N_f=4$ theory, colliding of the
roots $e_5=e_1=e_2=-m_1$ and $e_6=e_1=e_2=-m_1$ result into two
branches of parabola
\be
Y_1^2=(2m_1+m_2)^2 - \Lambda(2m_5+m_2)+ \Lambda^2=0
\label{par1}
\ee
in the mass $(m_1,m_2)$-plane, see fig.~\ref{fi:parw}. At $m_5 \gg
m_{1,2}$, after renormalization $\Lambda m_5\to
\Lambda_4^2$ one comes back to the straight lines from \rf{AD1}, but
generally one find from \rf{par1} that the size of the strong
coupled domain around the classical line $2m_1+m_2=0$ from
fig.~\ref{fi:lines} is rather $\sqrt{\Lambda\Delta
m}=\sqrt{\Lambda(m_5-m_1)}$ in $N_f=5$ theory (if we fix $\Delta
m$), which can be seen, rewriting \rf{par1} as
\be
\left(m_1+{m_2\over 2}-{\Lambda\over 2}\right)^2 = \Lambda(m_5-m_1)
\label{par11}
\ee
Like in $N_f =4$ case on one branch of this parabola the $r=2$ vacuum
collides with the $r=1$ vacuum, where quark $Q^2_2$ and monopole with charges
(\ref{1mnc}) are massless, while on the other branch it collides with another
$r=1$ vacuum with the massless quark $Q^2_2$ and dyon (\ref{a1dyc}).

\begin{figure}[tp]
\epsfysize=7cm
\centerline{\epsfbox{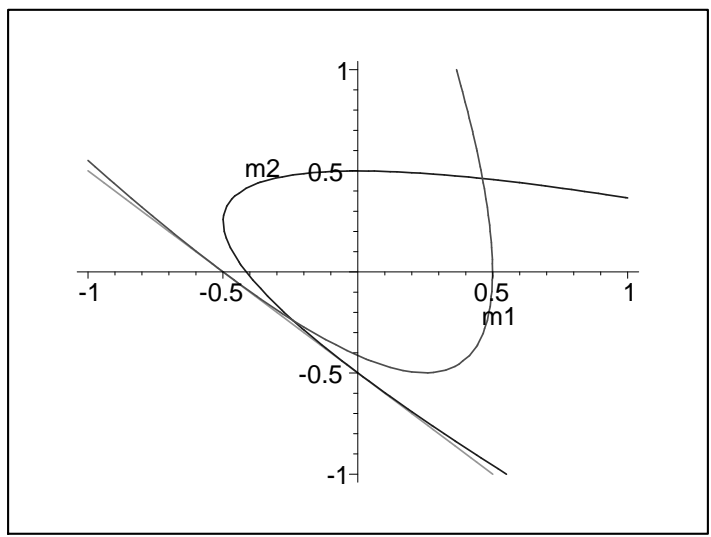}}
\caption{Strong-coupling parabolas in $N_f=5$ theory:
the lines \rf{par1} and \rf{par2} are plotted at $m_5=\Lambda$.}
\label{fi:pars}
\end{figure}
If instead we collide $e_5=e_3=e_4=-m_2$ and $e_6=e_3=e_4=-m_2$, one
gets the second parabola
\be
Y_2^2=(m_1+2m_2)^2 - \Lambda(2m_5+m_1)+ \Lambda^2=0
\label{par2}
\ee
or
\be
\left(m_2+{m_1\over 2}-{\Lambda\over 2}\right)^2 = \Lambda(m_5-m_2)
\label{par22}
\ee
with the same properties. On two branches of this parabola our $r=2$ vacuum collides with $r=1$
vacua with
massless quark $Q^1_1$ and the monopole (\ref{2mnc}) or dyon (\ref{a2dyc}) respectively.
The  intersection  of two parabolas is depicted at
fig.~\ref{fi:parw},~\ref{fi:pars} and we see how the rhombus of
$N_f=4$ theory is deformed in the $N_f=5$ case.
These results can be also confirmed by direct calculation of Seiberg-Witten
periods
\be
\label{Bkdg5}
a_k^D = {1\over 2\pi i}\oint_{B_k}dS =
 -{i\over \pi}\left(Y_k +\left(M+{m_k+m_5\over
2}\right)\log{M+m_k+\Lambda-Y_k\over M+m_k+\Lambda+Y_k} + \right.
\\
\left. +
{m_k-m_5\over 2}\log{a+b_+m_k-b_-Y_k\over a+b_+m_k+b_-Y_k}
\right),\ \ \ k=1,2
\\
a=(M-\Lambda)^2+m_5(M-3\Lambda),\ \ \ b_\pm=M+m_5\pm\Lambda
\ee
with $Y_{1,2}$ being defined in \rf{par1}, \rf{par2}. Formulas
\rf{Bkdg5} show that the conclusions about $N_f=4$ case remain
almost intact in the case of five flavors. At $Y_k=0$, $k=1,2$ (i.e.
for each parabola \rf{par1}, \rf{par2} in the mass plane) the
imaginary part of the corresponding period $\Im(a^D_k)$ vanishes,
while the real part $\Re(a^D_k)$ jumps when passing from the
positive to negative branch of the corresponding $k$-th parabola.

Permutation of roots upon crossing these AD-lines can be analyzed much in the same way as in
$N_f=4$ case. Say, on the first branch of the parabola (\ref{par11})
the Seiberg-Witten curve (\ref{su3nf5}) gives the equation
\beq
z(z^2-\epsilon z+\delta)=0
\eeq
which is the same as  eq.~(\ref{cu4e1}) (taken at the first AD line) with $\delta=2\sqrt{\Lambda(m_5-m_1)}\,\delta \phi$, while  $\epsilon$ is defined via relation
\beq
2m_1+m_2-\Lambda =2\sqrt{\Lambda(m_5-m_1)} +\epsilon.
\eeq
In fact, the scheme we
considered in previous section does not at all depend upon the
details of the model, and is completely described by the effective
$SU(2)$ curve with two cuts, which is considered in
Appendix~\ref{app:su2cu}.
Therefore the result of the root permutation is the same: inside the "deformed rhombus"
massless quarks $Q^1_1$ and $Q^2_2$ transform into massless dyons $D^1_1$ and $D^2_2$
with charges (\ref{dyon1}) and (\ref{dyon2}) respectively.

Hence, we come now already close to the main conclusion of our
paper. Decreasing the values of the fundamental masses, we turn the
original theory into the strong coupling regime. Due to monodromies
in the space of masses, the original light quarks in our model
change their quantum numbers and turn into the light dyons with the
charges \rf{dyon1} and \rf{dyon2}. The exact analysis of the
Seiberg-Witten theory shows that this happens, in particular, on the
AD lines, surrounding the strongly coupled domain in the mass space.
The exact shape of the strongly-coupled domain is slightly different
in the $N_f=4$ and $N_f=5$ theories, but that does not influence the
main conclusions. The $N_f=4$ theory is conformal, and therefore one
cannot really ensure for it the weakly coupled regime for the dual
theory of light dyons. However, in the $N_f=5$ case in the regime
\rf{semior}, inside the domain surrounded by parabolas on
fig.~\ref{fi:pars} (we shall still refer to it as to rhombus), the
effective dual theory of light dyons is at weak coupling, and the
semiclassical analysis of the properties of confinement \cite{MY} is
directly applicable. In order to do this, we are going to write the
effective action, and study the charges of the string solutions.

\section{Low-energy effective action at strong coupling}
\label{ss:dualact}
\setcounter{equation}{0}

In this section we construct the effective low-energy theory at
strong coupling, i.e. inside the rhombus on fig.~\ref{fi:dia_yu}, or
in the "rhombus" surrounded by two parabolas at fig.~\ref{fi:pars}.
We call it dual to the original theory with the action
(\ref{u2model}). In this region we keep $|m_A-m_B|$ small (see
(\ref{semior})) and $m\ll \Lambda$.

As was shown above, the   massless quarks $Q^1_A$ and
 $Q^2_A$ are transformed into the massless dyons $D^1_A$ and $D^2_A$ with the
 charges \rf{dyon1} and \rf{dyon2}; the latter
 form a fundamental
representation of the  gauge group $SU(2)$.  According to the
charges (\ref{dyon1}), \rf{dyon2}, the third component of the
$SU(2)$ dual gauge field has to be the following linear combination
\be
B^3_{\mu}= \frac1{\sqrt{2}}\,(A^{3}_{\mu}+A^{3D}_{\mu})
\label{B3}
\ee
of the gauge fields. If the dual non-Abelian gauge group is restored,
the components $B^{1,2}_{\mu}$ of the gauge field  become massless
at $m_1=m_2$. Let us check, whether this is indeed the case.

The electric and magnetic charges of the dual $W$-bosons $B_\mu^\pm
\sim B^{1}_{\mu}\mp iB^{2}_{\mu}$ coincide with the charges of the
operators $\tilde{D}^A_2 D^1_A$ and $\tilde{D}^A_1 D^2_A$. From
(\ref{dyon1}), \rf{dyon2} we get therefore for these charges
$(\bmu_1-\bmu_2)\oplus(\balpha_1-\balpha_2)=\balpha_{12}\oplus\balpha_{12}$,
see fig.~\ref{fi:su3}, or, in physical normalization, more
explicitly
\be
 B^{\pm}_{\mu}:\ \ \ \ \ {\bf n}_e = {\bf n}_m=
 \pm{\balpha_{12}\over\sqrt{2}},\ \ \ \
 {\rm or} \ \ \ \
\left(n_e^3,n_m^3;\,n_e^8,n_m^8\right)= \left(\,\pm1,\pm1;\,0,0\right)
\label{B12}
\ee
These charges determine the mass of these states by the
Seiberg--Witten mass formula \cite{SW1}. One gets from (\ref{mld})
\be
m_{B^\pm_{\mu}}=\sqrt{2}\left|a_3+ a^D_3\right|=\left|\Delta
m_{12}\right|
\label{WDmass}
\ee
At $\Delta m_{12}\to 0$, the dual W-boson mass \rf{WDmass} vanishes,
as was expected, and therefore the fields (\ref{B3}) and (\ref{B12})
can be unified into the adjoint
 multiplet $B^{\alpha}_{\mu}$, ($\alpha=1,2,3$), of the non-Abelian
 $SU(2)$ factor of the gauge group of the dual theory.

The light dyons $D^k_A$ ($k=1,2$) are also charged with respect to
the $U(1)$ gauge group, associated with the $T^8$-generator of the
underlying $SU(3)$ gauge group broken in the dual theory down to
$SU(2)\times U(1) \simeq U(2)$. According to the dyon charges
\rf{dyon1},
\rf{dyon2} the dual photon field is given by the following linear combination
\be
B^8_{\mu}= \frac1{\sqrt{10}}\,(A^{8}_{\mu}+3A^{8D}_{\mu})
\label{B8}
\ee
of the dual gauge fields. It turns out, that the dyons $D^k_A$ and
the gauge fields $B^{\alpha}_{\mu}$ ($\alpha=1,2,3$), $B^8_{\mu}$,
together with their superpartners, are the only light states to be
included in the dual low-energy effective theory inside the rhombus
of strong-coupled regime. All other states are either heavy (with
masses of the order of ${\tilde\Lambda}$ the scale \rf{tlam} of the
dual theory) or decay on the curves of marginal stability
\cite{SW1,SW2,BF,svz,SYcross,SYdual}.

It means, that the bosonic part of the effective low-energy action
of the dual theory inside the rhombus can be written in the form
\be
 S_{\rm dual} =\int d^4x \left[\frac{1}{4\tilde{g}^2_{2}}
\left(F^{\alpha}_{\mu\nu}\right)^2
 +\frac1{4\tilde{g}^2_1}\left(F^8_{\mu\nu}\right)^2
+\frac1{\tilde{g}^2_2}\left|D_{\mu}b^a\right|^2
\right.
\\
\left.
 +\frac1{\tilde{g}^2_8}
\left|\partial_{\mu}b^8\right|^2
+\left|\nabla_{\mu}
D^A\right|^2 + \left|\nabla_{\mu} \tilde{D}_A\right|^2
+
 V(D,\tilde{D},b^\alpha,b^8)\right]
\label{u2dual}
\ee
where $b^8$ and $b^\alpha$, so that
\be
b^8= \frac{1}{\sqrt{10}}\,(a^{8}+3a^{8}_D),\ \ \  b^3=
\frac{1}{\sqrt{2}}\,(a^{3}+a^{3}_D)\;\;\; {\rm for}\;\;\;
\alpha=3
\label{bbp}
\ee
are the scalar \ntwo superpartners of gauge fields $B^{8}_{\mu}$ and
$B^{\alpha}_{\mu}$, while $F_{\mu\nu}^8$ and $F_{\mu\nu}^\alpha$ are
their field strengths, and the gauge couplings $\tilde{g}_1$ and
$\tilde{g}_2$ correspond to the $U(1)$ and $SU(2)$ subgroups
respectively.

The covariant derivatives are defined in accordance with the charges
of the dyons, namely
\be
\nabla_\mu
=\pt_{\mu}-i\left(\sqrt{2}\,B^\alpha_{\mu}\frac{\tau^\alpha}{2}
 +\sqrt{\frac{5}{6}}\,B^8_{\mu}\right)
\label{nablaD}
\ee
The scalar potential $V(D,\tilde{D},b^\alpha,b^8)$ in the action
(\ref{u2dual}) is
\be
 V(D,\tilde{D},b^\alpha,b^8) =
 \frac{\tilde{g}^2_2}{4}
\left( \bar{D}^A\tau^\alpha D_A -
\tilde{D}^A \tau^\alpha \bar{\tilde{D}}_A \right)^2
+ \frac{5}{12}{\tilde g}^2_1
\left(|D_A|^2 -|\tilde{D}^A|^2 \right)^2 +
\\
+ \tilde{g}_2^2\left| \tilde{D}^A \tau^\alpha D_A + \frac{\pt {\cal
W}}{\pt b^\alpha}\right|^2 +\tilde{g}_1^2\left|
\sqrt{\frac{5}{3}}\tilde{D}^A D_A+
\frac{\pt {\cal W}}{\pt b^8}\right|^2 +
\\
+\sum_{A=1}^{N_f} \left[
\left|\left(\sqrt{\frac{5}{3}}b^8 +\tau^\alpha b^\alpha + m_A\right)D_A
\right|^2
+
\left|\left(\sqrt{\frac{5}{3}}b^8 +\tau^\alpha b^\alpha+m_A\right)
\bar{\tilde{D}}_A
\right|^2 \right]
\label{dualpot}
\ee
Now let us turn directly to the desired limit of equal quark masses,
$\Delta m_{AB}=0$. The vacuum of the theory (\ref{u2dual}) is
located, due to \rf{mld},
\rf{bbp}, at the following values of the scalar condensates
\be
\langle b^8\rangle=-\sqrt{\frac35}{m_1+m_2\over 2}=-\sqrt{\frac35}m,
\ \ \ \
\langle b^\alpha \rangle= 0
\label{bvev}
\ee
while the VEV's of the dyon fields are determined by the FI F-term
coming from the derivatives of \none deformation superpotential
${\cal W}$ in (\ref{dualpot}). For these derivatives one can write
from general principles
\be
\frac{\pt {\cal W}}{\pt b^8}={\tilde\mu}\Lambda+\ldots,
\ \ \ \ \
\frac{\pt {\cal W}}{\pt b^\alpha}=c{\tilde\mu}b^\alpha +\ldots
\ee
where $ \tilde{\mu}={\rm const }\mu$ and dots stand for higher powers of fields $b^8$ and $b^\alpha$ we
ignore at small ${\tilde\mu}$, while the (inessential for our
purposes) constant $c$ can be in principle determined from the exact
solution. This gives
\be
\langle D^k_A\rangle =\langle \tilde{D}_k^A\rangle=\sqrt{\frac{\tilde{\xi}}{2}}
\underbrace{\left(
\begin{array}{ccccc}
1 &  0 & 0 & 0 & 0\\
0 &  1 & 0 & 0 & 0\\
\end{array}
\right)}_{N_f}
\label{Dvev}
\ee
where we restrict ourselves to the case $N_f=5$, while
\be
\tilde{\xi}= \sqrt{\frac35}{\tilde\mu}{\Lambda}
\label{tildexi}
\ee
One can also calculate the dimension of the Higgs branch which
emerges in the equal mass limit. It results in
\be
 {\rm dim} {\cal H}\,\Big|_{\rm rhombus}= 4N N_f  - 2N^2 -N^2 -N^2
=4N (N_f-N)\Big|_{N=2}=8(N_f-2)
\label{dimHdual}
\ee
where we have to take into account the $4N N_f $ real dyon degrees
of freedom and subtract $2N^2$ $F$-term conditions, $N^2$ $D$-term
conditions and, finally, $N^2$ phases eaten by the Higgs mechanism,
where $N=2$ for the $U(2)$ dual gauge group. The dimension of the
Higgs branch  of the dual theory \rf{dimHdual} coincides, as
expected, with the dimension of the Higgs branch \rf{dimH} in the
original low energy theory (\ref{u2model}).

From (\ref{bvev}) and (\ref{Dvev}) we see again that both gauge
$U(2)$ and flavor $SU(N_f)$ groups, are broken in the vacuum, but
the color-flavor locked form of (\ref{Dvev}) guarantees that the
diagonal global $SU(2)_{C+F}$ survives. More exactly, the unbroken
global group of the dual theory is
$SU(2)_{C+F}\times SU(N_f-2)_F\times  U(1)$
and coincides again with the global group of the original theory
(for generic quark masses is broken down to $U(1)^{N_f-1}$). Much in
the same way as in the original  theory (\ref{u2model}), the
presence of the global $SU(2)_{C+F}$ group leads to formation of the
non-Abelian strings in the dual theory \rf{u2dual}.

\section{Confined monopoles
\label{ss:conf}}
\setcounter{equation}{0}

Since quarks are in the Higgs phase in the weak coupling regime of
the original theory at large $m$ in $r=2$ vacuum, the monopoles are
confined. Two of three $SU(3)$ elementary monopoles with the charges
$\balpha_{1,2}$, or $(\,0,\pm\frac12;\,0,\frac{\sqrt{3}}{2})$, see
\rf{1mnc} and \rf{2mnc}, are attached to the ends of the elementary strings while the
third one with the charge $\balpha_{12}=\balpha_1-\balpha_2$ or
$(\,0,1;\,0,0)$ becomes a string junction of two elementary strings
\cite{MY}. Inside the rhombus at small $m$ the dual theory
(\ref{u2dual}) is in the weak coupling regime and we can use it to
study confinement at strong coupling. In this domain the light dyons
condense instead quarks, therefore, here we deal with oblique
confinement \cite{thooft}.

In this section we determine the elementary string fluxes of the
strings in the dual theory (\ref{u2dual}) inside the rhombus
 and show that still the elementary monopole
fluxes can be absorbed by  strings. Hence, it is the {\em monopoles}
being still confined,  much in the same way as in the original
$U(2)$ theory (\ref{u2model}) at weak coupling.

Consider, first, the elementary  string $S_1$ arising due to the
winding of dyon $D^1_1$. At $r\to\infty$ (in the transverse plane
with polar co-ordinates $(r,\theta)$ to the direction of the string)
one has
\be
\left. D^1_1\right|_{r\to\infty} \sim\sqrt{\tilde{\xi}}\,e^{i\theta},
\ \ \ \
\left. D^2_2\right|_{r\to\infty} \sim \sqrt{\tilde{\xi}}
\label{Dwind}
\ee
see (\ref{Dvev}). Taking into account the dyon charges
(\ref{dyon1}), \rf{dyon2} we derive the behavior of the gauge
potentials at infinity,
\be
{\bf n}^1\cdot({\bf A}_i\oplus{\bf A}^D_i) =
{1\over\sqrt{2}}\left(\bmu_1\cdot{\bf A}_i +\balpha_1\cdot{\bf
A}^D_i \right)=
 \frac{A_i^3}{2} +\frac{A_i^{3D}}{2} + \frac{ A_i^8}{2\sqrt{3}}
+ \frac{\sqrt{3}}{2} A_i^{8D}\sim \pt_i \theta
\\
{\bf n}^2\cdot({\bf A}_i\oplus{\bf A}^D_i) =
{1\over\sqrt{2}}\left(\bmu_2\cdot{\bf A}_i +\balpha_2\cdot{\bf
A}^D_i
\right)= -\frac{A_i^3}{2} -\frac{A_i^{3D}}{2}+ \frac{A_i^8}{2\sqrt{3}} +
\frac{\sqrt{3}}{2} A_i^{8D} \sim 0
\ee
which, in turn, implies both
\be
 A_i^3 + A_i^{3D}  \sim  \,\pt_i \theta
\\
   \frac{A_i^8}{\sqrt{3}}
+ \sqrt{3} A_i^{8D} \sim \,\pt_i \theta
\label{gaugewindint}
\ee
The combinations orthogonal to those of (\ref{gaugewindint}) are
required to vanish at infinity: $A_i^3 - A_i^{3D}\sim 0$ and
$A_i^{8D} -3A_i^{8} \sim 0$. As a result one gets for each component
\be
 A_i^3  \sim \frac12 \,\pt_i \theta, \ \ \ \ \ A_i^{3D}  \sim \frac12 \,\pt_i \theta\,,
\\
  A_i^8 \sim \frac{\sqrt{3}}{10}\,\pt_i \theta,\ \ \ \ \
 A_i^{8D} \sim \frac{3\sqrt{3}}{10}\,\pt_i \theta
\label{gaugewindrhombus}
\ee
The string charges are defined in terms of the fluxes
\be
\oint dx_i (A^{3D}_i,A^3_i;A^{8D}_i,A^8_i)= 4\pi\,(-n^3_e,\,n^3_m;-n^8_e,\,n^8_m)\,.
\label{defstrch}
\ee
This definition ensures that the string has the same charge as a
trial dyon which can be attached to the string endpoint (not
necessarily being present in the spectrum of the theory). In
particular, according to this definition, the charge of the
$S_1$-string in dual theory with the fluxes (\ref{gaugewindrhombus})
is
\be
{\bf n}_{S_1}=
\left(\,-\frac14,\,\frac14;\,-\frac{3\sqrt{3}}{20},\,\frac{\sqrt{3}}{20}\right)
\label{S1}
\ee
In a similar way one determines the charges of another
$\mathbb{Z}_2$-elementary string, arising due to winding at spatial
infinity of the second dyon $D^2_2$:
\be
{\bf n}_{S_2}=
\left(\,\frac14,\,-\frac14;\,-\frac{3\sqrt{3}}{20},\,\frac{\sqrt{3}}{20}\right)
\label{S2}
\ee
Now we can check that, each of three $SU(3)$ monopoles can be indeed
confined by these two strings. Say, for the monopoles with the
charges ${\bf 0}\oplus\balpha_{1,2}$ or
$(\,0,\pm\frac12;\,0,\frac{\sqrt{3}}{2})$ one has
\be
{1\over\sqrt{2}}({\bf 0}\oplus\balpha_1) =
(\,0,\frac12;\,0,\frac{\sqrt{3}}{2})= {\bf n}_{S_1} +\frac7{10} {\bf
n}^1+\frac2{10}{\bf n}^2,
\\
{1\over\sqrt{2}}({\bf 0}\oplus\balpha_2) =
(\,0,-\frac12;\,0,\frac{\sqrt{3}}{2})= {\bf n}_{S_2} +\frac2{10}
{\bf n}^1+\frac7{10}{\bf n}^2
\label{confrhombus}
\ee
where ${\bf n}^1$ and ${\bf n}^2$ are the charges of the $D^1_1$ and
$D^2_2$ dyons (\ref{dyon1}), \rf{dyon2}. Formula \rf{confrhombus}
shows, that only a part of the monopole flux is confined to the
string, while the remaining part is screened by dyon condensate.
Finally, for the third $SU(3)$ $\balpha_{12}$-monopole one gets from
(\ref{confrhombus}), that
\be
{1\over\sqrt{2}}({\bf 0}\oplus\balpha_{12}) = (0,1;\,0,0)= {\bf
n}_{S_1} - {\bf n}_{S_2} +\frac12\left( {\bf n}^1-{\bf n}^2\right)
\label{conf12}
\ee
and we find that it is also confined, being a junction of two
elementary strings $S_1$ and $S_2$.

We see that although the quark charges change as we pass from the
weak to strong coupling domains, and they become dyons, this does
{\em not} happen to the monopoles: the monopole states do not
change their charges, and they are confined by elementary strings in
both domains at large and small $m$. However, inside the rhombus in
dual theory there is a peculiarity: the monopole flux is only
partially carried by attached string, and the remaining part of it
is screened by the dyon condensate.

Hence, our result provides an explicit counterexample to the
commonly accepted belief, that if monopoles are confined in the
original theory, then  quarks should be confined in the dual
theory. We have demonstrated above however, that it is monopoles
monopoles rather than quarks are confined in the strong coupling
domain at small $m$. Similar results are obtained in \cite{SYdual}
for \ntwo supersymmetric $U(N)$ QCD with number of flavors $N_f<2N$.

\section{Conclusions
\label{ss:concl}}
\setcounter{equation}{0}

In this paper we have answered to the long-standing question, what
happens at strong coupling to the confinement of monopoles in
supersymmetric QCD, caused by the flux tubes in the
effective theory around $r=2$ quark vacua \cite{MY}. In order to do
this, the picture from the original theory at large
values of quark masses has been moved into the domain of small masses,
or strongly
coupled original theory. Fortunately, using the exact Seiberg-Witten
solution to the \ntwo supersymmetric gauge theories, one can go
beyond the semiclassical approximation and study at least, what
happens to the quantum numbers of the light fields.

We have considered in this paper the supersymmetric QCD with large
number of flavors (masses some of them were taken coinciding for
simplicity) and demonstrated, that when going towards the domain of
small masses the light quarks acquire magnetic charges. This can be
seen by careful study of   permutation of the
branch points and the period integrals on (almost) singular curves in the vicinity of
colliding vacua. As a result, in the domain of small masses one
deals with the effective theory of light dyons.

When masses of the original condensed quarks coincide, the
non-Abelian gauge symmetry is partially restored. The exact form of
the Seiberg-Witten curves shows, that the same phenomenon occurs in
the dual effective theory of dyons, i.e. at equal values of
masses it becomes non-Abelian. As a consequence, the string or flux
tube solutions in the effective theory acquire the non-Abelian
structure, quite in the same way, as in the original theory, and
moreover, the spectrum of the confined objects is again non-Abelian.

The charges of confined objects are determined by
string fluxes and charges of the condensates. The analysis of
the last section shows, that despite the condensate in the dual
theory is formed by dyons and its charge contains the magnetic component, the
charges of the confined objects, i.e. those parts of string's
fluxes, not being screened by the condensate, are still {\em
magnetic}, and {\em not electric}. The generic  reason
for this is distinction between the weight-quark charges, and
the root-monopole ones. Coming to the strongly coupled domain, a
quark can acquire the monopole charge, but cannot get rid of the
original electric weight-vector. Hence, the quarks cannot be
confined  by strings, formed due to the condensation of dyons,
which possess the quark electric charges in addition to the
magnetic ones: the states, confined by these strings, are still monopoles.

Of course, there are $r=0$ vacua in our theory where the monopoles (or dyons
with root electric charges) condense, and electric flux tubes are formed. This
triggers confinement of quarks according to Seiberg-Witten scenario \cite{SW1}.
However, this provides a model with only Abelian confinement. As we have shown in this paper,
if being interested in the non-Abelian confinement, which occurs due to the
formation of non-Abelian strings, we ultimately end up with the confinement of
monopoles. This conclusion is in accordance with similar results, obtained
in \cite{SYdual} for the \ntwo supersymmetric $U(N)$ theory with $N_f<2N$.

Although \ntwo supersymmetric QCD does not looks like the real world QCD,
the supersymmetric gauge theories can be considered as certain
"theoretical laboratory" for studying the properties of realistic
confinement. One can be really amused, how
the strongly coupled region can be described in terms of the dual
theory, and the properties of all objects we are interested in are
under the full control. Hence, studying the different phases of
supersymmetric QCD provides an important experience - what can in
principle happen in gauge theory at strong coupling, and is
therefore very useful not only in the context of mathematical
physics, but enriching our intuition before attacking the problems
of the real world.

\section*{Acknowledgments}

One of us (AY) is grateful to A. Gorsky and M. Shifman for useful
discussions.

The work of AM was supported by the Russian Federal Nuclear Energy
Agency, by Grant of Support for Scientific
Schools LSS-1615.2008.2, by the RFBR grants 08-01-00667,
09-02-90493-Ukr, 09-02-93105-CNRSL, 09-01-92440-CE, by Kyoto
University, and by the Dynasty Foundation. The work of
AY was supported by FTPI, University of Minnesota, by the RFBR Grant
09-02-00457a and by Grant of Support for Scientific Schools
LSS-11242003.2. AM thanks IHES at
Bures-sur-Yvette, where a part of this work had been done, and
the Yukawa Institute for Theoretical Physics, where the work
has been completed.


\section*{Appendix}
\appendix
\setcounter{equation}0

\section{Classical picture in different gauges
\label{app:gauge}}

Here we present for completeness the explicit pictures for the
condensate charges in two different gauges. In addition to the gauge
\rf{fir2} we use throughout the main text, we relate it to the "global" or
"real" gauge, where all eigenvalues of $\Phi = {\rm
diag}(\phi_1,\phi_2,\phi_3)$ are chosen to be real and ordered, say
$\phi_1<\phi_2<\phi_3$. We always take the 12-vacuum, with condensed
two first flavors with the bare masses $m_A$, $A=1,2$.

\begin{figure}[tp]
\epsfysize=8cm
\centerline{\epsfbox{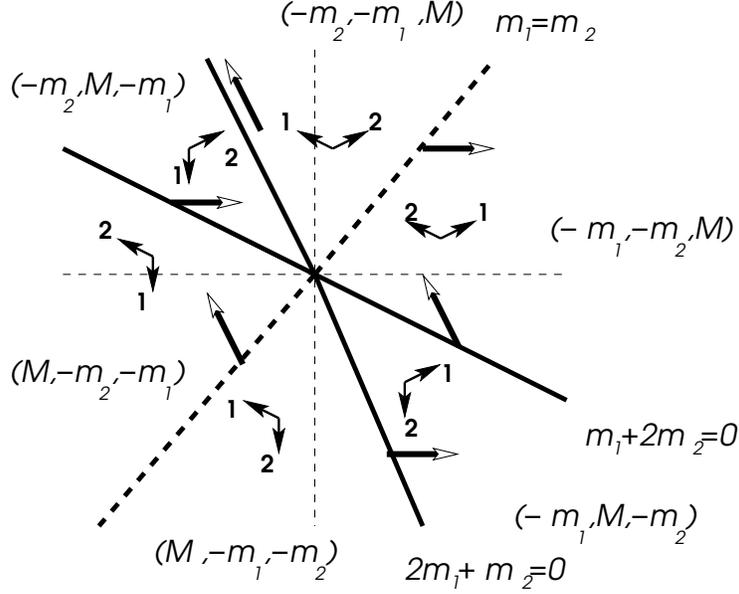}}
\caption{Classical picture for the 12-vacuum in the mass $(m_1,m_2)$-plane
($M\equiv m_1+m_2$),
for the gauge, when
the eigenvalues of $\Phi$ are real and ordered
$\phi_1<\phi_2<\phi_3$. The diagonal part of the $\Phi$-matrix at
each corner, and the direction of each $A=1,2$ condensed flavor are
presented explicitly. The long root vector at each line shows also
the color-direction of each restored $SU(2)$-subgroup in this
gauge.}
\label{fi:lines3}
\end{figure}

The directions of the condensed flavors in color space are depicted
in the global gauge directly at each sector, as well as the form of
the matrix $\Phi$, which is different from \rf{fir2} just by
particular permutation of the eigenvalues. On the straight lines one
gets a restoration of $SU(2)$-subgroup, again in each case its
direction in the color space is given by a root, presented
explicitly at fig.~\ref{fi:lines3}.

The dashed line differs from the solid ones by the fact that both
quarks on each side of it are charged w.r.t. the restored $SU(2)$,
and passing through this line just exchanges the colors of two
condensed flavors, which can be seen in semiclassical regime. In
contrast to that, on each solid line, only one of the quark flavors
is charged w.r.t. restored $SU(2)$ group, and it flips it direction
in the color space, while the other one is remained intact. This
process goes in regime beyond the semiclassical approximation, and
is governed by the Seiberg-Witten mechanism \cite{SW1,SW2}. In the
first case, on the dashed lines, flipping of the quark's charge is
caused by the massless $(1,0)$ W-boson in the restored $SU(2)$,
while on the dashed lines it is caused by the massless $(1,1)=[2,1]$
dyon.

\begin{figure}[tp]
\epsfysize=8cm
\centerline{\epsfbox{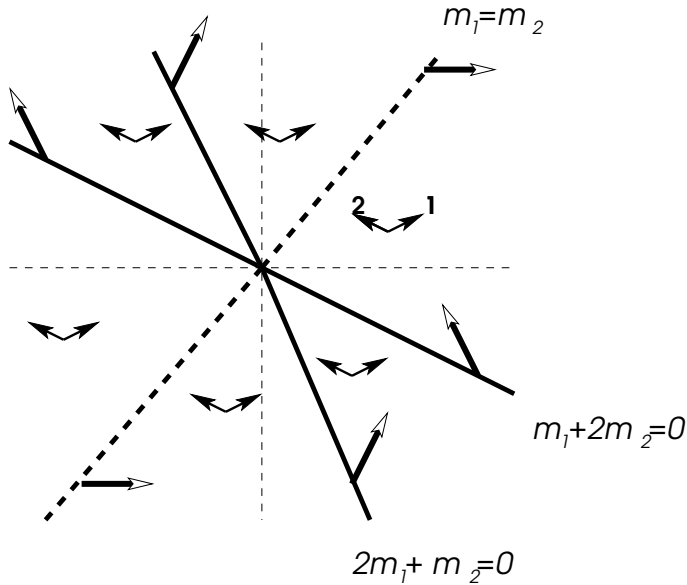}}
\caption{Classical picture for the 12-vacuum in the mass $(m_1,m_2)$-plane
for the gauge, when matrix $\Phi$ is always of the form
\rf{fir2}. The direction of each
$A=1,2$ condensed flavor then remains the same in each sector, while
the color-direction of each restored $SU(2)$-subgroup are rotated
correspondingly.}
\label{fi:lines12}
\end{figure}
The picture from fig.~\ref{fi:lines3} can be gauge transformed to
fix the matrix $\Phi$ being always of the form \rf{fir2}, in the
whole $(m_1,m_2)$-plane. The result is presented on
fig.~\ref{fi:lines12}, and we see that all vectors in the color
space are just rotated, consistently for the weights of quarks and
roots of the restored $SU(2)$ subgroups. For the sake of simplicity
(e.g. all the restored $SU(2)$-subgroups have the same color
direction on different branches of the same solid line) we use the
gauge \rf{fir2} and that of fig.~\ref{fi:lines12} in the main text,
though the consistency between the different sectors of the
weakly-coupled theory and the charges of Seiberg-Witten dyons is
better seen in the global gauge, i.e. at fig.~\ref{fi:lines3}.

\section{Period integrals on degenerate curves
\label{app:deg}}


In the basic example of $N_c=2$, $N_f=2$ theory with the coinciding
masses $m_1=m_2=m$ one gets for \rf{cuy}
\be
\label{n2n2}
y^2 = \left( x^2-u\right)^2 - 4\Lambda^2(x+m_1)(x+m_2) =
\left( x^2-u\right)^2 - 4\Lambda^2(x+m)^2
\ee
and the generating differential \rf{dS} turns for the pairwise
coinciding masses into
\be
dS \sim {x dP\over y} - x {P\over 2y}{dQ\over Q}+ \frac12 x{dQ\over
Q} = {x dP\over y} - x {P\over y}{dq\over q}+ x{dq\over q}
\\
q(x) = \prod_{B=1}^{N_f/2}(x+m_B)
\label{dSm2}
\ee
where we have chosen $m_{B+N_f/2}=m_B$, $B=1,\ldots,N_f/2$ for even
number of flavors $N_f$. The curve \rf{n2n2} just corresponds to
$P(x)=x^2-u$ and $q(x)=x+m$.

When exactly at quark vacuum $u=u_Q=m^2$, the curve \rf{n2n2}
degenerates further to
\be
y^2 = (x+m)^2\left((x-m)^2-4\Lambda^2\right)
\equiv (x+m)^2Y^2
\\
Y^2= (x-m)^2-4\Lambda^2
\label{n2n2q}
\ee
and the Seiberg-Witten differential \rf{dS} turns into
\be
dS = {xdx\over Y}+{xdx\over x+m}
\label{dSm2q}
\ee
Due to \rf{resdS} and to the fact, that on degenerate curve the
position of the mass pole at $x=-m$ (for both flavors) coincides
with the degenerate cut, the differential \rf{dSm2q} is normalized
by\footnote{Here again the ``homological'' normalization
of the charges and periods is used, so that the quark's quantum numbers are
$(n_e,0)=(\half,0)=[1,0]$. To get the ``physical normalization'' ${\sf a}$, one
should just renormalize in \rf{resdSm2q} (and in \rf{reaDm} below) $a=
n_e\cdot {\sf a} = \half {\sf a}$. }
\be
{1\over 2\pi i}\oint_{x=-m} dS_+ = {1\over 2\pi i}\oint_{A^+} dS_+ =
a = -m
\\
{1\over 2\pi i}\oint_{x=-m} dS_- = \res_{x=-m}dS_- - {1\over 2\pi
i}\oint_{A^-} dS_- = -2m+ a = -m
\label{resdSm2q}
\ee
obviously true for \rf{dSm2q}. Since the curve
\rf{n2n2q} is rational, the differential
\rf{dSm2q} can be easily integrated, giving rise to
\be
S = Y + m\log(x-m+Y)+ x -m\log(x+m)
\label{Sm2q}
\ee
In order to compute the desired $B$-period (the monopole mass), one
has to take the difference $\left.S_+\right|_{x=-m} -
\left.S_-\right|_{x=-m}$ of the values of \rf{Sm2q} on two different
sheets of the Riemann surface \rf{n2n2q}.

This is not possible to do by direct substitution of $x=-m$ into
\rf{Sm2q} due to the logarithmic singularity, i.e. the curve
\rf{n2n2q} is "too degenerate". Let us then regularize it and denote
the distance between the position of the pole and the nearest end of
the shrinking cut by $\epsilon^\pm$, dependently on the sheet
$Y=Y_\pm$ of \rf{n2n2q}. The values of these
$\epsilon^\pm=\epsilon^\pm(m,\Lambda)$ can be determined as follows
(see e.g. \cite{MN}): the differential
\be
d\phi = {dw\over w}\ \stackreb{\rf{n2n2q}}{=}\ {dx\over Y}+{dx\over
x+m}
\label{dphi}
\ee
should have constant periods \cite{KriW}, moreover, its $B$-periods
on \rf{n2n2q} can be just chosen vanishing. Integrating \rf{dphi} up
to
\be
\phi = \log (x-m+Y) + \log (x+m)
\label{phi}
\ee
and putting $\left.\phi_+\right|_{x=-m} -
\left.\phi_-\right|_{x=-m} \equiv \left.\phi_+\right|_{x=-m+\epsilon^+} -
\left.\phi_-\right|_{x=-m+\epsilon^-}=0$, one gets (at
$\epsilon^\pm\to 0$)
\be
\log{\epsilon^+\over\epsilon^-} =
\log {m+\sqrt{m^2-\Lambda^2}\over m-\sqrt{m^2-\Lambda^2}}
\label{eml}
\ee
Therefore
\be
\left.S_+\right|_{x=-m} - \left.S_-\right|_{x=-m} =
-m\log{\epsilon^+\over\epsilon^-} + 2\left.Y\right|_{x=-m} +
m\log{-2m + \left.Y\right|_{x=-m}\over -2m - \left.Y\right|_{x=-m}}
=
\\
= 4\sqrt{m^2-\Lambda^2}+  2m\log {m-\sqrt{m^2-\Lambda^2}\over
m+\sqrt{m^2-\Lambda^2}}
\label{SD}
\ee
Hence, evaluating $B$-period on degenerated curve
\rf{n2n2q} gives rise to explicit formula (cf. the result with \cite{Dorey,HaHo})
\be
a_D = {1\over 2\pi i}\left(\left.S_+\right|_{x=-m} -
\left.S_-\right|_{x=-m}\right) = -{i\over
\pi}\left(2\sqrt{m^2-\Lambda^2}+m
\log{m-\sqrt{m^2-\Lambda^2}\over m+\sqrt{m^2-\Lambda^2}}\right)
\label{aDm}
\ee
showing, that $\left.\Im(a_D)\right|_{m=\pm \Lambda}=0$. However,
one should also consider carefully the real part of \rf{aDm}, taking
into account the logarithmic cut. We fix it to vanish at
$m=\Lambda$, then
\be
\left.\Re (a_D)\right|_{m=\Lambda}=0
\\
\left.\Re (a_D)\right|_{m=-\Lambda} = 2m = -2a = -{\sf a}
\label{reaDm}
\ee
It means, that when quark singularity $u_Q$ collides with $u_M$ we
have massless monopole with $|a_D|=0$, while when $u_Q$ collides
with $u_D$, one gets the vanishing mass of the $(1,1)=[2,1]$ dyon,
$|a_D+2a|=|a_D+{\sf a}|=0$. Quite in a similar way, using the degenerate curves \rf{nf4deg},
\rf{nf5dg}, \rf{u1n1} and corresponding limit of the Seiberg-Witten differential
\rf{dS} one computes the periods \rf{B1dg}, \rf{B2dg} and \rf{Bkdg5}.

\section{Permutation of the branching points on $SU(2)$ curve
\label{app:su2cu}}

\begin{figure}[tp]
\epsfysize=7cm
\centerline{\epsfbox{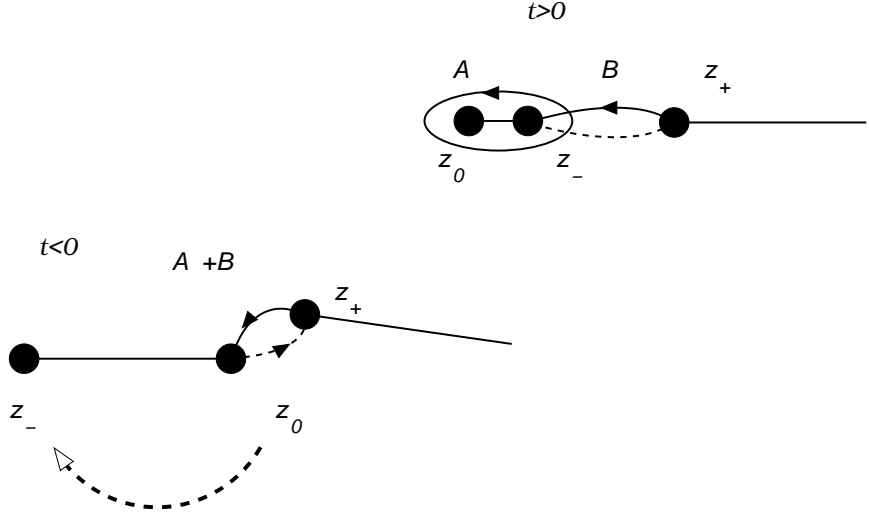}}
\caption{Permutation of the branch points at $t\sim 0$. The
$A$-cycle "catches" the $B$-cycle and turns into $A+B$, which means
that  massless $[1,0]$ quark emits the massless
anti-monopole and turns into the massless dyon $\tilde{a}+m+a^D\to 0$.}
\label{fi:z1}
\end{figure}
Let us now turn to the issue of permutation of the branching points
of the "basic" Seiberg-Witten curve \rf{n2n2} in the vicinity of its
degenerate form \rf{n2n2q}. Obviously one can rewrite \rf{n2n2} as
\be
y^2 = [x^2-u-2\Lambda(x+m)][x^2-u+2\Lambda(x+m)] = \\
\\ =
[z^2-2(m+\Lambda)z+\delta][z^2-2(m-\Lambda)z+\delta]
\label{n2n2tr}
\ee
where the co-ordinate $x=-m+z$ is just shifted, and the distance
$\delta$ is introduced
\be
u = m^2 - \delta,\ \ \ \delta >0
\label{del}
\ee
so that for $\delta\to 0$ one gets back to
\rf{n2n2q}. Introducing also
\be
t = m-\Lambda
\label{tlm}
\ee
the curve \rf{n2n2tr} acquires the form
\be
y^2 = [z^2-2(2\Lambda+t)z+\delta][z^2-2tz+\delta]\
\stackreb{|t|\ll\Lambda}{\simeq}\ [z^2-4\Lambda z+\delta][z^2-2tz+\delta]
\label{n2n2td}
\ee
convenient to study around the monopole singularity, where the
r.h.s. reflects what happens, when we approach the region
$m\sim\Lambda$ and $t\ll
\Lambda$. The roots of \rf{n2n2td} are at
\be
z_0 \simeq {\delta\over 4\Lambda} \approx 0, \ \ \ z_\Lambda \simeq 4\Lambda\to\infty\\
{\rm and}
\\
z_\pm = t \pm \sqrt{t^2-\delta}
\label{rt}
\ee
If $t^2 \gg \delta$ and $t$ is real positive, the obvious ordering
is
\be
(z_0 \simeq {\delta\over 4\Lambda} \approx 0) < (z_- \simeq
{\delta\over 2t}) < (z_+ \simeq 2t) < (z_\Lambda
\simeq 4\Lambda \approx \infty)
\label{odro}
\ee
so that the natural choice for the $A$-cycle is around $z_0$ and
$z_-$, while for the dual $B$-cycle from $z_-$ to $z_+$, see
fig.~\ref{fi:z1}.  At positive $t$
the $A$-cycle, drawn as a contour around $z_0$ and
$z_-$ at fig.~\ref{fi:z1}, is ``small'' i.e. almost shrinks.
However at negative $t$, the roots
$z_+$ and $z_0$ become close to each other, and the shrinking contour
is now $A+B$, cf. with sect.~\ref{ss:pbp}. Thus, the quark picks up the monopole charge and
transforms into the massless
dyon with the mass $|a+a_D +m|=|\half{\sf a}+a_D+m|=0$.

\begin{figure}[tp]
\epsfysize=7.5cm
\centerline{\epsfbox{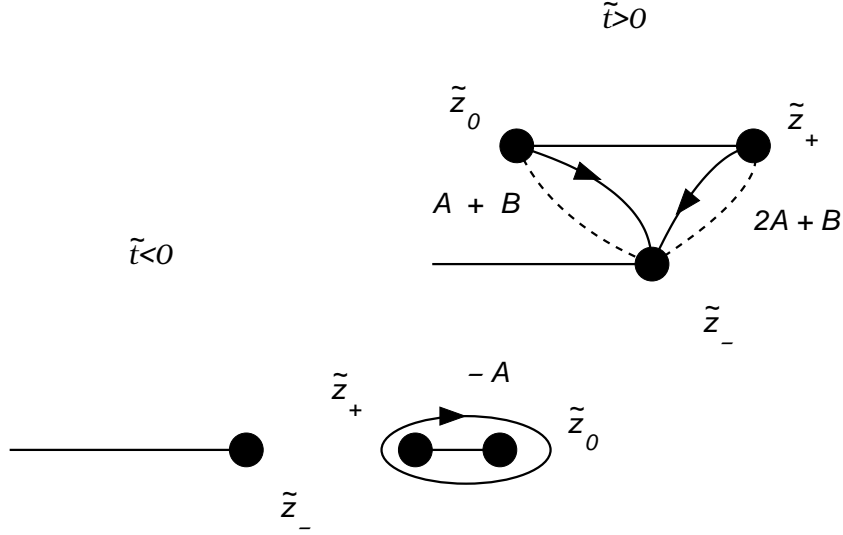}}
\caption{Exchange of roots in the regime ${\tilde t}\sim 0$. The "small" cycle $A+B$ (the upper
picture), encircling ${\tilde z}_0$ and ${\tilde z}_-$, is combined
with the degenerate $2A+B$ at ${\tilde z_+} = {\tilde z}_-$ (taken
with the opposite sign), and they form again the ``quark cycle''.}
\label{fi:z2}
\end{figure}
To study another interesting domain one needs to introduce
\be
{\tilde t} = m+\Lambda
\label{ttlm}
\ee
instead of \rf{tlm} and consider the curve \rf{n2n2tr} for ${\tilde
t} \ll \Lambda$, i.e. as
\be
y^2 = [z^2-2{\tilde t}z+\delta][z^2-2({\tilde t}-4\Lambda)z+\delta]\
\stackreb{|{\tilde t}|\ll\Lambda}{\simeq}\ [z^2-2{\tilde t}z+\delta][z^2+4\Lambda z+\delta]
\label{n2n2ttd}
\ee
which is basically just a result of replacement $t\to
4\Lambda-{\tilde t}$. Instead of
\rf{rt} one gets the roots
\be
\label{subst12}
z_+ \mapsto {\tilde z}_0\simeq -{\delta\over 4\Lambda} \approx 0,
\ \ \ \
z_- \mapsto {\tilde z}_\Lambda\approx -4\Lambda
\\
z_0\simeq {\delta\over 4\Lambda}\approx 0 \mapsto {\tilde z}_-,
\ \ \ \
z_\Lambda\approx 4\Lambda \mapsto {\tilde z}_+
\label{rtt}
\ee
where
\be
{\tilde z}_\pm = {\tilde t} \pm
\sqrt{{\tilde t}^2-\delta}
\label{tzpm}
\ee
The "small" $A+B$ cycle is now surrounding the points $z_+ \mapsto
{\tilde z}_0$ and $z_0\mapsto {\tilde z}_-$, and at ${\tilde t}\gg
\sqrt{\delta}$ we have the natural ordering
\be
({\tilde z}_\Lambda \simeq -4\Lambda \approx -\infty) < ({\tilde
z}_0 \simeq -{\delta\over 4\Lambda} \approx 0) < ({\tilde z}_-
\simeq {\delta\over 2{\tilde t}}) < ({\tilde z}_+ \simeq 2{\tilde t})
\label{odrot}
\ee
As we reduce  ${\tilde t}$ the dyon cycle $2A+B$
becomes also degenerate. At negative ${\tilde t}$
the roots
${\tilde z}_0$ and ${\tilde z}_+$ become close to each other, and
the contour $(A+B)-(2A+B)=-A$ almost shrinks, see fig.~\ref{fi:z2}. This gives again
a quark cycle, corresponding to the  massless quark.



\begin{thebibliography}{7799}

\bibitem{MY}
 A.~Marshakov and A.~Yung,
  Nucl. Phys. {\bf B647} (2002) 3
  [arXiv:hep-th/0202172].

\bibitem{SW1}
N.~Seiberg and E.~Witten, Nucl. Phys. {\bf B426} (1994) 19,
[arXiv:hep-th/9407087].

\bibitem{SW2}
N.~Seiberg and E.~Witten, Nucl. Phys. {\bf B431} (1994) 484,
[arXiv:hep-th/9408099].

\bibitem{APS}
P.~Argyres, M.~Plesser and N.~Seiberg, Nucl. Phys. {\bf B471} (1996)
159, [arXiv:hep-th/9603042].

\bibitem{CKM}
G.~Carlino, K.~Konishi and H.~Murayama, Nucl. Phys. {\bf B590}
(2000) 137, [arXiv:hep-th/0005076].

\bibitem{HT1}
A.~Hanany and D.~Tong,
JHEP {\bf 0307}, 037 (2003) [arXiv:hep-th/0306150].

\bibitem{ABEKY}
R.~Auzzi, S.~Bolognesi, J.~Evslin, K.~Konishi and A.~Yung,
Nucl. Phys. {\bf B673}, 187 (2003) [arXiv:hep-th/0307287].

 \bibitem{SYmon}
M.~Shifman and A.~Yung,
Phys. Rev.  {\bf D70}, 045004 (2004) [arXiv:hep-th/0403149].

\bibitem{HT2}
A.~Hanany and D.~Tong,
JHEP {\bf 0404}, 066 (2004) [arXiv:hep-th/0403158].

\bibitem{Trev}
D.~Tong,
{\em TASI Lectures on Solitons,}
  arXiv:hep-th/0509216.

\bibitem{Jrev}
  M.~Eto, Y.~Isozumi, M.~Nitta, K.~Ohashi and N.~Sakai,
  J.\ Phys.   {\bf A 39}, R315 (2006)
  [arXiv:hep-th/0602170].

\bibitem{SYrev}
M.~Shifman and A.~Yung,
Rev.\ Mod.\ Phys. {\bf 79} 1139 (2007)
[arXiv:hep-th/0703267];
{\sl Supersymmetric Solitons,}
Cambridge University Press, 2009.


\bibitem{Trev2}
D.~Tong
{\em Quantum Vortex Strings: A Review,}
arXiv:0809.5060 [hep-th].

\bibitem{SYdual}
M.~Shifman and A.~Yung,
Phys. Rev. {\bf  D79}, 125012 (2009)
[arXiv:0904.1035 [hep-th]].


\bibitem{SYcross}
M.~Shifman and A.~Yung,
Phys. Rev. {\bf D 79}, 105006 (2009)
  arXiv:0901.4144 [hep-th].

\bibitem{Sdual}
  N.~Seiberg,
  Nucl.\ Phys.\  {\bf B435}, 129 (1995)
  [arXiv:hep-th/9411149].

\bibitem{IS}
K.~A.~Intriligator and N.~Seiberg,
  Nucl.\ Phys.\ Proc.\ Suppl.\  {\bf 45BC}, 1 (1996)
  [arXiv:hep-th/9509066].

\bibitem{HSZ}
A.~Hanany, M.~J.~Strassler and A.~Zaffaroni,
Nucl. Phys. {\bf B513}, 87 (1998)
[hep-th/9707244].

\bibitem{VY}
A.~Vainshtein and A.~Yung,
Nucl. Phys. {\bf B614}, 3 (2001) [arXiv:hep-th/0012250].


\bibitem{MMY}
V.~Markov, A.~Marshakov and A.~Yung,
  Nucl. Phys. {\bf B709} (2005) 267
  [arXiv:hep-th/0408235].

\bibitem{AchVas}
A.~Achucarro and T.~Vachaspati,
  Phys.\ Rept.\  {\bf 327}, 347 (2000)
  [arXiv:hep-ph/9904229].

\bibitem{SYsem}
 M.~Shifman and A.~Yung,
  Phys. Rev.  {\bf D73}, 125012 (2006)
  [arXiv:hep-th/0603134].

\bibitem{Jsem}
M.~Eto, J.~Evslin, K.~Konishi, G.~Marmorini, M.~Nitta, K.~Ohashi, W.~Vinci, N.~Yokoi,
  Phys. Rev.  {\bf D76}, 105002 (2007)
  [arXiv:0704.2218 [hep-th]].

\bibitem{EY}
 K.~Evlampiev and A.~Yung,
 Nucl. Phys. {\bf B662} (2003) 120
 [hep-th/0303047].


\bibitem{Khrev} N.Dorey, T.Hollowood, V.Khoze and M.Mattis, Phys.Rept. {\bf 371}
(2002) 231-459, arXiv:hep-th/0206063

\bibitem{KlMa}
 T.~Grimm, A.~Klemm, M.~Marino and M.~Weiss,
  JHEP {\bf 0708} (2007) 058
  [arXiv:hep-th/0702187].

\bibitem{Zam_agt}
 A.~Marshakov, A.~Mironov and A.~Morozov,
  JHEP {\bf 0911} (2009) 048
  [arXiv:0909.3338 [hep-th]].

\bibitem{HO} A.~Hanany and Y.~Oz,
  Nucl. Phys. {\bf B452} (1995) 283
  [arXiv:hep-th/9505075].

\bibitem{ArFa}
P. C.~Argyres and A. E.~Faraggi,
Phys. Rev. Lett. {\bf 74},  3931 (1995) [arXiv:hep-th/9411057].

\bibitem{KLTY}
A.~Klemm, W.~Lerche, S.~Yankielowicz and S.~Theisen,
Phys. Lett. {\bf B344}, 169 (1995) [arXiv:hep-th/9411048].

\bibitem{ArPlSh}
P.~Argyres, M.~Plesser, and  A.~Shapere,
Phys. Rev. Lett. {\bf 75}, 1699 (1995)
[arXiv:hep-th/9505100].


\bibitem{GMMM} A.Gorsky, A.Marshakov, A.Mironov and A.Morozov,
Phys.Lett., {\bf B380} (1996) 75-80, [arXiv:hep-th/9603140].


\bibitem{LMN}
A.Losev, A.Marshakov and N.Nekrasov, in Ian Kogan memorial volume
{\it From fields to strings: circumnavigating theoretical physics},
581-621; [arXiv:hep-th/0302191].

\bibitem{MN} A.Marshakov and N.Nekrasov,
  JHEP {\bf 0701} (2007) 104, hep-th/0612019;\\
  A.Marshakov,
  Theor.Math.Phys.  {\bf 154} (2008) 362
  [arXiv:0706.2857[hep-th]].

\bibitem{AD}
P.~C.~Argyres and M.~R.~Douglas,
Nucl. Phys. {\bf B448}, 93 (1995) [arXiv:hep-th/9505062].

\bibitem{APSW}
P.~Argyres, M.~Plesser, N.~Seiberg, and E.~Witten,
Nucl. Phys. {\bf B461}, 71 (1996)
[arXiv:hep-th/9511154].


\bibitem{BF}
A.~Bilal and F.~Ferrari,
Nucl. Phys. {\bf B516}, 175 (1998) [arXiv:hep-th/9706145].

\bibitem{svz}
M.~Shifman, A.~Vainshtein and R.~Zwicky,
J. Phys.  {\bf A 39}, 13005 (2006)
hep-th/0602004.

\bibitem{thooft}
G. 't Hooft, in {\it 1981 Carg\'{e}se Summer School Lecture Notes on
Fundamental Interactions, NATO Adv. Study Inst. Series B: Phys.,}
Vol. 85, ed. M. L\'{e}vy {\it et al.}  (Plenum Press, New York, 1982)
[reprinted in G. 't Hooft, {\it Under the Spell of the Gauge Principle}
(World Scientific, Singapore, 1994), page 514];
Nucl. Phys.  {\bf B190},  455 (1981);
see also
J.~Cardy and E.~Rabinovici, Nucl. Phys.  {\bf B205}, 1 (1982);
J.~Cardy, Nucl. Phys. {\bf B205}, 17 (1982).


\bibitem{KriW}
I.~Krichever,
Commun. Pure. Appl. Math. {\bf 47} (1992) 437 [arXiv:
hep-th/9205110].

\bibitem{Dorey}
 N.~Dorey,
  JHEP {\bf 9811} (1998) 005
  [arXiv:hep-th/9806056];
\\
 N.~Dorey, T.~Hollowood and D.~Tong,
  JHEP {\bf 9905} (1999) 006
  [arXiv:hep-th/9902134].

\bibitem{HaHo}
A.~Hanany and K.~Hori,
  Nucl. Phys. {\bf B513}, 119 (1998)
  [arXiv:hep-th/9707192].






\end{thebibliography}
\end{document}